\documentclass[a4paper,11pt]{article}
\pdfoutput=1
\usepackage{paperstyle}

\usepackage{amsmath,amssymb,graphicx}
\usepackage{soul}
\usepackage[latin1]{inputenc}
\usepackage{dsfont}

\usepackage{subfigure}

\usepackage{array}
\usepackage{mathtools}
\usepackage{bbold}

\usepackage[thinlines]{easytable}

\newcommand{\be}{\begin{eqnarray}}
\newcommand{\en}{\end{eqnarray}}

\newcommand{\Hubble}{ H }

\newcommand{\mpl}{m_{\rm{Pl}}}
\newcommand{\As}{A_{\rm{s}}}
\newcommand{\ns}{n_{\rm{s}}}

\newcommand{\braket}[1]{\langle #1 \rangle}

\def\low{\rule{0pt}{2.0ex}}
\def\Low{\rule{0pt}{1.5ex}}

\usepackage[section, below, above]{placeins}

\usepackage{afterpage}

\begin{document}

\begin{titlepage}

\vspace*{-15mm}
\vspace*{0.7cm}

\begin{center}

{\Large {\bf Parametric resonance after hilltop inflation caused by an\\[1mm]inhomogeneous inflaton field}}\\[8mm]

Stefan Antusch$^{\star\dagger}$\footnote{Email: \texttt{stefan.antusch@unibas.ch}},  
Francesco Cefal\`{a}$^{\star}$\footnote{Email: \texttt{f.cefala@unibas.ch}},
David Nolde$^{\star}$\footnote{Email: \texttt{david.nolde@unibas.ch}} and 
Stefano Orani$^{\star}$\footnote{Email: \texttt{stefano.orani@unibas.ch}}

\end{center}

\vspace*{0.20cm}

\centerline{$^{\star}$ \it
Department of Physics, University of Basel,}
\centerline{\it
Klingelbergstr.\ 82, CH-4056 Basel, Switzerland}

\vspace*{0.4cm}

\centerline{$^{\dagger}$ \it
Max-Planck-Institut f\"ur Physik (Werner-Heisenberg-Institut),}
\centerline{\it
F\"ohringer Ring 6, D-80805 M\"unchen, Germany}

\vspace*{1.2cm}

\begin{abstract}
\noindent We study preheating after hilltop inflation where the inflaton couples to another scalar field, e.g.\ a right-handed sneutrino, which provides a mechanism for generating the correct initial conditions for inflation and also a decay channel for the inflaton that allows for reheating and non-thermal leptogenesis. In the presence of such a coupling, we find that after the phases of tachyonic preheating and tachyonic oscillations, during which the inflaton field becomes inhomogeneous, there can be a subsequent preheating phase where the fluctuations of the other field get resonantly enhanced, from initial vacuum fluctuations up to amplitudes of the same order (and even larger) as the ones of the inflaton field. This resonant enhancement differs from the usual parametric resonance as the inflaton field is inhomogeneous at the time the enhancement takes place. We study this effect using lattice simulations as well as semi-analytically with a generalized Floquet analysis for inhomogeneous background fields.

\end{abstract}
\end{titlepage}

\tableofcontents

\section{Introduction}

Cosmic inflation, a phase of accelerated expansion of the early universe, has various attractive features. It can solve the flatness and horizon problems if it lasts sufficiently long, i.e.\ more than about 60 $e$-folds, and it can provide the seeds for structure formation in excellent accordance with CMB observations \cite{Aghanim:2015xee,Ade:2015lrj,Ade:2015tva} and galaxy surveys \cite{Beutler:2011hx,Anderson:2013zyy,Ross:2014qpa}, provided that the inflaton field's potential has an appropriate shape. Currently, still a variety of inflation models is in agreement with observations, although ongoing and future experiments will narrow down the set of viable models.

Towards a more fundamental theory of the early universe, another task has to be approached: inflation has to be embedded into a full theory of particle physics. This is necessary in order to investigate the reheating of the universe after inflation, i.e.\ the transition from potential energy domination to radiation domination and thermal equilibrium \cite{Abbott:1982hn,Albrecht:1982mp,Traschen:1990sw,Kofman:1994rk,Khlebnikov:1996mc,Khlebnikov:1996zt,Kofman:1997yn,Davidson:2000er,Felder:2000hr,Micha:2004bv,Allahverdi:2010xz,Amin:2014eta}. To study this transition, one has to know how the inflaton field interacts with the other fields of the theory in order to finally produce the particles our universe consists of today. 

Among the viable models of inflation, small-field hilltop inflation \cite{Linde:1981mu,Izawa:1996dv,Izawa:1997df,Senoguz:2004ky} allows for particularly interesting connections to particle physics, as the inflationary dynamics can be associated with the spontaneous breaking of a symmetry at energies near the GUT scale. Indeed, as recently shown in \cite{Antusch:2014qqa}, the couplings to other scalar fields can dynamically generate the initial conditions of the inflaton close to the top of the hill, necessary to get inflation started, via a phase of ``preinflation''.\footnote{For further discussions related to the the initial conditions for hilltop inflation, see e.g.\ \cite{Izawa:1997df,Vilenkin:1983xq,Vilenkin:1994pv,Boubekeur:2005zm}.} Furthermore, they provide the decay channels for reheating and may generate the baryon asymmetry of the universe via leptogenesis. A particularly interesting scalar field in this respect is the right-handed sneutrino, the superpartner of one of the right-handed neutrinos predicted in the supersymmetric version of the seesaw mechanism to explain the observed neutrino masses. It can initially act as the ``preinflaton'' to create suitable initial conditions for inflation to start, and, if produced in sufficient abundance after inflation, can explain the baryon asymmetry of the universe via CP violating out-of-equilibrium decays into pairs of Higgs(inos) and (s)leptons. To calculate quantities like the non-thermally produced baryon asymmetry or the abundance of cosmic relics like gravitinos,\footnote{For non-thermal gravitino production in a model with similar field content, see \cite{Antusch:2015tha}.} it is important to understand the dynamics of reheating after inflation.

In this paper, we study preheating after hilltop inflation. In particular, we consider a coupling between the inflaton $\phi$ and another scalar field $\chi$ allowing for a ``preinflation'' mechanism as described in \cite{Antusch:2014qqa}. It is known that after hilltop inflation, the inflaton field can very efficiently become inhomogeneous via the processes of tachyonic preheating and tachyonic oscillations \cite{Felder:2000hj,Desroche:2005yt,Brax:2010ai,Antusch:2015nla}, where, throughout this paper, by inhomogeneous we mean that the energy density is dominated by sub-Hubble perturbations with $k/a > H$. We will see that considering an additional scalar field allows for a third phase of preheating: a parametric resonance of $\chi$ caused by the inhomogeneous inflaton field $\phi$. It is known that parametric resonance can be a very effective mechanism of particle production. However, it is usually caused by a homogeneous field, whereas in this case the inflaton field is inhomogeneous at the time the resonance takes place.\footnote{Amplification of other scalar fields' perturbations from inhomogeneous background fields during preheating has already been observed in \cite{Felder:2000hr}, though the authors did not study its parametric dependence and did not interpret it as a parametric resonance. Also, parametric resonance has been observed during planar domain wall collisions \cite{Braden:2014cra} which constitute a different kind of inhomogeneous background.} We study this process using lattice simulations as well as semi-analytically with a generalized Floquet analysis for inhomogeneous background fields. We find the band structure of the resonance and show that the fluctuations of $\chi$ can be excited from the initial vacuum state up to amplitudes of the order of the inflaton's fluctuations, leading to comparable particle abundances.

The paper is organized as follows: the hilltop inflation model studied in this paper is introduced in section~\ref{sec:hilltopModel}. In section~\ref{sec:preheating}, we briefly discuss the dynamics of preheating after hilltop inflation. Subsequently, we study the parametric resonance of an additional scalar field $\chi$ which is coupled to the inhomogeneous inflaton field. To this end, we perform a semi-analytical generalized Floquet analysis in section~\ref{sec:floquet}. For more accurate results, we performed numerical lattice simulations which are presented and discussed in section~\ref{sec:num_analysis_lattice}. Finally, we summarize and conclude in section~\ref{sec:conclusions}.

\section{Hilltop inflation with coupling to another scalar field}
\label{sec:hilltopModel}

In this paper, we consider the hilltop inflation potential
\begin{align}
 V(\phi, \chi) \, = \, V_0 \left( 1 - \frac{\phi^6}{v^6} \right)^2 + \frac{\lambda^2}{2}\phi^2\chi^2(\phi^2+\chi^2)\,,
\label{eq:potential}
\end{align}
with $v \ll \mpl$. The coupling $\lambda$ has units of $\mpl^{-1}$ and couples the inflaton to another scalar field $\chi$.  

Inflation happens while the inflaton $\phi$ slowly rolls away from the maximum ($\phi = 0$) and $\chi \sim 0$. As we can always perform a field redefinition $\phi \rightarrow \pm\phi$, we choose $\phi>0$ during inflation. The slow-roll regime ends when $\eta_{\phi}\equiv m^2_{\rm{Pl}}(\partial^2V/\partial\phi^2)/V\simeq-1$, and $\phi$ quickly rolls towards the minimum $\phi = v$.

It will turn out that it is useful to define the masses of the fields at the minimum of the potential ($\phi=v$ and $\chi=0$):
 \begin{align}
m^2_{\phi} \, \equiv \, \left.\frac{\partial^2 V}{\partial \phi^2}\right|_{\rm{min}} \, = \,\frac{72 V_0}{v^2}\,, 
\label{eq:def_mass_phi}
\end{align}
\begin{align}
m^2_{\chi} \, \equiv \, \left.\frac{\partial^2 V}{\partial \chi^2}\right|_{\rm{min}} \, = \,\lambda^2\,v^4\,.
\label{eq:def_mass_chi}
\end{align}
Accordingly, the mass ratio is related to the coupling constant $\lambda$ via
\begin{align}
\frac{m_{\chi}}{m_{\phi}} \, = \, \frac{\lambda\,v^3}{\sqrt{72\,V_0}}\,.
\label{eq:massratio-lambda-relation}
\end{align}

\begin{figure}
\centering
\includegraphics[width=13cm]{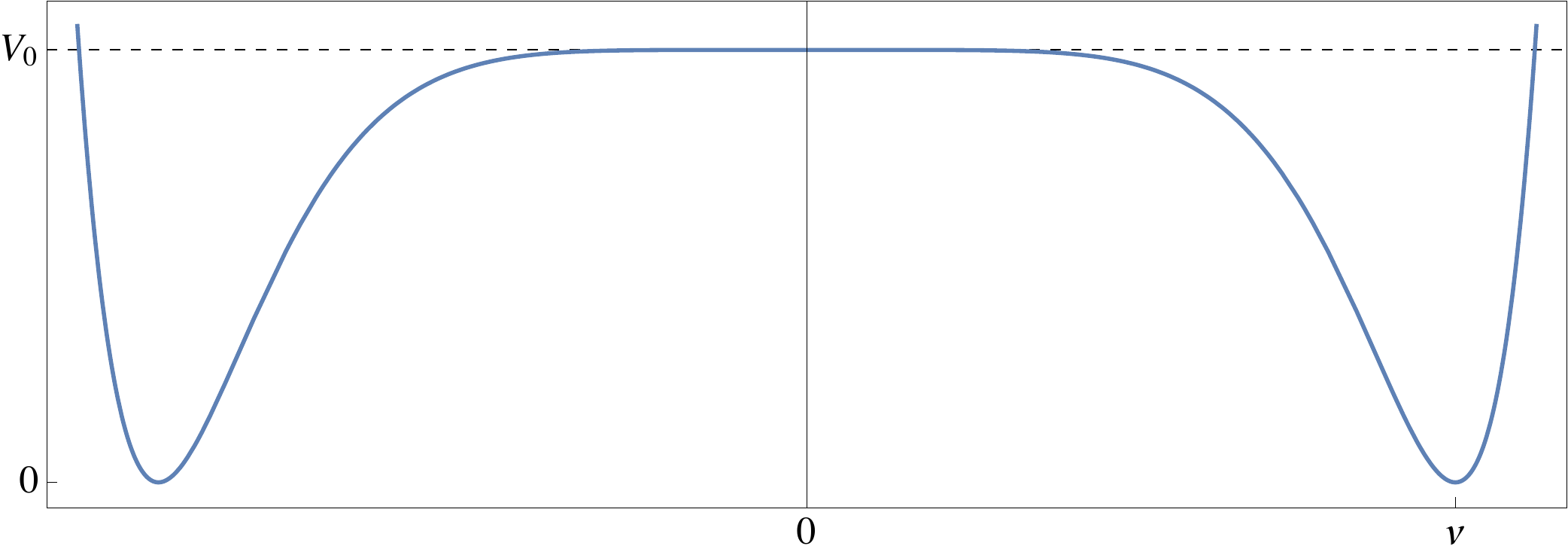}
\caption{Illustration of the inflaton potential eq.~\eqref{eq:effective_potential_inflation}. Inflation happens near $\phi\approx 0$, while the inflaton slowly rolls towards its minimum $\phi=v$. }
\label{fig:potential}
\end{figure}
\subsection*{Initial conditions and preinflation}
The initial conditions for inflation ($\phi\approx0$) can be generated dynamically by a stage of preinflation \cite{Izawa:1997df,Senoguz:2004ky,Yamaguchi:2004tn,Antusch:2014qqa}. If $\chi$ is non-zero, a $\chi$ dependent mass is generated for $\phi$ which stabilizes the inflaton near the hilltop ($\phi = 0$). When $\chi$ falls below a sufficiently small value, the inflaton destabilizes due to quantum fluctuations and the system evolves towards one of the global minima $\phi = \pm v$ and $\chi = 0$. The latter phase can be described as single-field hilltop inflation with the effective potential 
\begin{align}
V(\phi) \, &= \,V_0 \left( 1 - \frac{\phi^6}{v^6} \right)^2\,,
\label{eq:effective_potential_inflation}
\end{align}
where the slow-roll phase lasts more than the required $N_* \sim  60$ $e$-folds to solve the flatness and horizon problems. An illustration of the potential $V(\phi)$ is shown in fig.~\ref{fig:potential}.

Depending on the model parameters, preinflation can also end with $\braket{\chi} \neq 0$. In sections~\ref{sec:floquet} and \ref{sec:num_analysis_lattice}, we study preheating with an initial $\braket{\chi}=0$ to get a conservative estimate of the minimal production of $\chi$ from vacuum fluctuations. The effects of an initial $\braket{\chi} \neq 0$ from preinflation are discussed afterwards in appendix~\ref{sec:nonzerochi}.

\subsection*{Predictions for the primordial perturbations}
Inflation ends at the inflaton field value $\phi_{\rm{e}}/m_{\rm Pl}\simeq((v/m_{\rm Pl})^6/60)^{1/4}$ when $\eta(\phi_{\rm{e}})\simeq-1$. The predictions for the primordial spectrum are calculated when the relevant perturbations exit the horizon, which is typically at about $N_* \sim 50$ -- $60$ $e$-folds before the end of inflation. The field value at horizon crossing, $\phi=\phi_*$, can be calculated from the number of $e$-folds
\begin{align}
N_* \, = \,\frac{1}{m^2_{\rm{Pl}}}\int^{\phi_*}_{\phi_{\rm{e}}}\frac{V\,d\phi}{\partial V/\partial\phi}\,,
\label{eq:Def_numb_efolds}
\end{align} 
by solving for $\phi_*$. The inflationary observables are related to the slow-roll parameters $\eta_{\phi}$ and $\varepsilon_{\phi}\equiv \frac{1}{2}m^2_{\rm{Pl}}(\partial V/\partial\phi)^2/V^2$, evaluated at $\phi=\phi_*$. This yields the predictions for the spectral index $\ns$ and the tensor-to-scalar ratio $r$:
\begin{align}
\ns \, &\simeq \, 1-6\,\varepsilon_{\phi}(\phi_*)+ 2\,\eta_{\phi}(\phi_*) \, \simeq \, 1-\frac{10}{5+4\,N_*}\, \simeq \,0.96,\\ 
r\, &\simeq \, 16\,\varepsilon_{\phi}(\phi_*)\, \simeq \,4\times10^{-6}\,\left(\frac{v}{m_{\rm{Pl}}}\right)^3\,,
\label{eq:ns_and_r}
\end{align}
which are compatible with the most recent Planck bounds $\ns = 0.9655\pm0.0062$ \cite{Aghanim:2015xee}. The value of $V_0$ can be deduced from the observed value of the scalar amplitude ($\As\simeq2.2\times10^{-9}$):
\begin{align}
V_0 \, = \, 24\pi^2\varepsilon_\phi({\phi_*})\As\,m_{\rm Pl}^4 \, \simeq \, 10^{-13}\,v^3\,m_{\rm{Pl}}\,.
\label{eq:V0}
\end{align}

\subsection*{Supersymmetric formulation}
Within a supersymmetric framework, the scalar potential~\eqref{eq:potential} can be constructed from the superpotential
\begin{align}
W \, = \, \sqrt{V_0}\,S\left(1-\frac{8\,\Phi^6}{v^6} \right) + \lambda\Phi^2X^2\,, 
\label{eq:superpotential}
\end{align}
where the fields have been promoted to chiral superfields.\footnote{We use the same symbol for a chiral superfield and its scalar component.} We assume that $S$ gets a super-Hubble mass during inflation from K\"{a}hler corrections (cf.\ \cite{Antusch:2008pn}) which confines it at zero during this phase. We will not include $S$ in the discussion of preheating in the main text, but discuss it separately in appendix \ref{appendix:S}. With $S=0$, the scalar potential is given by:
\begin{align}
V(\phi,\chi) \, &= \, \left|\frac{\partial W}{\partial S} \right|_{\theta=0}^2 + \left|\frac{\partial W}{\partial\Phi} \right|_{\theta=0}^2 + \left|\frac{\partial W}{\partial X} \right|_{\theta=0}^2 + V_{\rm{SUGRA}}\nonumber\\
\, &= \, V_0 \left( 1 - \frac{\phi^6}{v^6} \right)^2 + \frac{\lambda^2}{2}\phi^2\chi^2(\phi^2+\chi^2) + V_{\rm{SUGRA}} + \dots\,,
\label{eq:scalar_potential_sugra}
\end{align}
where $\phi = \sqrt{2}\,\rm{Re}[\Phi]$, $\chi=\sqrt{2}\,{\rm Re}[X]$ and the dots stand for terms containing the imaginary components of $\Phi$ and $X$. The imaginary inflaton component can in principle affect the dynamics and lead to different model predictions for the primordial spectrum \cite{Nolde:2013bha}. For most of the parameter space, however, the model reduces to the single-field limit. In this paper we therefore consider $\rm{Im}[\Phi]=0$. Furthermore, the potential~\eqref{eq:scalar_potential_sugra} depends only on $|X|$ and thus in the same way on ${\rm Re}[X]$ and ${\rm Im}[X]$. Since both fields have the same equation of motion, we only include $\chi=\sqrt{2}\,{\rm Re}[X]$ in our numerical lattice simulations. 

$V_{\rm{SUGRA}}$ contains additional K\"{a}hler corrections that contribute to the masses of $\phi$ and $\chi$. These corrections can be small or negligible either accidentally or due to a symmetry of the respective fields in the K\"{a}hler potential (e.g.\ a Heisenberg symmetry \cite{Antusch:2013eca}). For the purpose of this paper, we assume that these supergravity corrections are negligible so that the scalar potential is well-described by eq.~\eqref{eq:potential}. 

In a supersymmetric framework, one can easily fix the form of the superpotential by imposing charges for the superfields. The superpotential~\eqref{eq:superpotential}, for example, results directly from imposing a $U(1)_{\rm{R}}$ and a $\mathds{Z}_6$ symmetry if one distributes two units of $U(1)_{\rm{R}}$ charge to $S$, one unit to $X$, and zero to $\Phi$, and two units of $\mathds{Z}_6$ charge to $X$ and one unit to $\Phi$ while $S$ is uncharged under $\mathds{Z}_6$.

A possible candidate for the superfield $X$ is e.g.\ a right-handed sneutrino superfield, which provides a mechanism for generating the correct initial conditions for inflation and also a decay channel for the inflaton that allows for non-thermal leptogenesis \cite{Antusch:2014qqa}.

\section{Preheating dynamics}
\label{sec:preheating}
 
In this section we briefly discuss the phases of preheating for the hilltop inflation model~\eqref{eq:potential}. For what follows it is useful to consider the Fourier decomposition $f(\vec{x},t)\equiv\int\!\frac{d^3k}{(2\pi)^3}f_{\vec{k}}(t)e^{i\vec{k}\vec{x}}$ of the fields $f\in\{\phi,\chi\}$.
When the coupling constant $\lambda$ is zero, we can generically distinguish between two phases \cite{Brax:2010ai,Antusch:2015nla}:

\begin{itemize}

\item[{\textbf I}] \textbf{Tachyonic preheating}: at the end of inflation, the inflaton $\phi$ rolls down the potential towards $\phi=v$. Tachyonic amplification leads to exponential growth of all $\phi_{\vec{k}}$ for which $k/a < \sqrt{-\partial^2 V/\partial\phi^2}$. This growth is most efficient for very small $v$. For $v\lesssim10^{-5}m_{\rm{Pl}}$, tachyonic preheating amplifies the initial vacuum fluctuations to values $\langle\delta\phi^2\rangle \gtrsim v^2$, so the perturbations already become non-linear during this phase \cite{Brax:2010ai}.

\item[{\textbf II}] \textbf{Tachyonic oscillations}: $\phi$ oscillates around $\phi=v$, periodically entering the tachyonic region when $\partial^2 V/\partial\phi^2<0$ at values $\phi<(5/11)^{1/6}v$. Tachyonic amplification and damping of the oscillations due to the expansion of the universe lead to growth of the $\phi_{\vec{k}}$ around $k_{\rm peak} \sim 300 H_{\rm{i}}\equiv 300\sqrt{V_0/3}\simeq 0.2m_{\phi}$. For large $v\gtrsim10^{-1}m_{\rm{Pl}}$, however, the oscillations are strongly damped due to Hubble friction. As a consequence, the amplitude strongly decreases and $\phi$ quickly leaves the tachyonic region and never re-enters. In this case, preheating is not very efficient and the perturbations of $\phi$ remain linear. On the other hand, for $10^{-5}\,m_{\rm{Pl}} < v < 10^{-1}\,m_{\rm{Pl}}$, the fluctuations eventually become so large that in some regions the system develops localized bubbles which oscillate between the two minima $\phi=\pm v$. These oscillons are typically separated by a distance $\lambda_{\rm{peak}}\sim2\pi/k_{\rm{peak}}$ \cite{Antusch:2015nla}.
\end{itemize}
Including a coupling to an additional scalar field $\chi$ can potentially change the dynamics during preheating. For the model~\eqref{eq:potential}, we will see that for certain values of $\lambda$ one can have a third phase of preheating, during which the fluctuations of $\chi$ experience a strong growth:

\begin{itemize}

\item[{\textbf III}] \textbf{Parametric resonance of $\chi$ due to inhomogeneous inflaton background}: the dynamics of the $\phi$ fluctuations during and after the tachyonic oscillations can trigger a resonant amplification of $\chi$ fluctuations. The latter displays the characteristic features of a parametric resonance, since the way this phase occurs strongly depends on the value of the coupling constant $\lambda$ (or equivalently on the ratio $m_{\chi}/m_{\phi}$). 

\end{itemize}

In what follows, we will study this third phase of preheating after hilltop inflation. We are particularly interested in the case where the resonant amplification of $\chi$ fluctuations is sourced by an inhomogeneous inflaton background. As explained above, for $10^{-5}\,m_{\rm{Pl}} < v < 10^{-1}\,m_{\rm{Pl}}$, the inflaton becomes inhomogeneous during the first few oscillations around its minimum $\phi=v$.\footnote{The effect of the coupling between the inflaton and $\chi$ on the oscillons that form during the tachyonic oscillations (see \cite{Antusch:2015nla}) are studied in \cite{Antusch:2015ziz}.} For the present paper, we will consider the case $v=10^{-2}\,m_{\rm{Pl}}$.

In section~\ref{sec:floquet}, we apply the Floquet theory to study parametric resonance due to an inhomogeneous $\phi$ background under simplifying assumptions. For a more realistic view of the non-linear stage of preheating we performed numerical lattice simulations. The results are presented in section~\ref{sec:num_analysis_lattice}.

\section{Generalized Floquet analysis for parametric resonance with inhomogeneous background}
\label{sec:floquet}

When a homogeneous inflaton field $\phi$ oscillates around the minimum of its potential, this generates a nearly periodic variation in the masses of the fields coupled to the inflaton, e.g.\ the $\chi$ field in our model. Such a time-dependent mass can lead to efficient non-perturbative production of $\chi$ particles. This effect is called ``parametric resonance'' \cite{Kofman:1994rk,Kofman:1997yn,Allahverdi:2010xz,Amin:2014eta}.

In our hilltop model, the situation is different. During the phase of tachyonic oscillations, the homogeneous mode of $\phi$ decays into perturbations peaked around the characteristic scale $k_{\rm peak}$ \cite{Brax:2010ai,Antusch:2015nla}.\footnote{At least for $v \sim 10^{-2}\mpl$, which is the case considered in this paper.} A priori, it is not clear whether parametric resonance for $\chi$ can also work in this case: the usual mode equations do not apply, because they assume a homogeneous background field $\phi(t)$, but the $\chi$ field gets a locally time-dependent mass term due to the time evolution of $\phi(\vec{x},t)$ and could therefore be non-perturbatively amplified.

In this section, we want to understand analytically how $\chi$ might undergo a parametric resonance even for an inhomogeneous inflaton field whose perturbations are peaked at some $k = k_{\rm peak} \gg aH$. To make the problem tractable, we use four simplifying assumptions:
\begin{enumerate}
 \item We assume that $\delta \phi := (\phi - v) \ll v$. This assumption becomes valid briefly after the end of tachyonic oscillations, as Hubble damping quickly reduces the amplitude of perturbations $\delta \phi$.\footnote{This assumption makes our analysis inapplicable during the earlier stage of preheating, when $\delta \phi \sim v$. We must refer to our lattice simulations in section~\ref{sec:num_analysis_lattice} to decide whether $\chi$ can grow non-perturbatively during this early phase.}
 \item We assume $\delta\chi \ll \delta \phi$. This assumption is valid initially, as $\delta\chi$ is not strongly amplified during the homogeneous oscillations of $\phi$. It can be violated later if there is efficient $\delta\chi$ production due to preheating. In that case, our formalism breaks down at the time when $\delta\chi \sim \delta \phi$.
 \item We approximate $\delta\phi(\vec{x},t)$ as a standing wave with wavenumber $\lvert \vec{k} \rvert = k_{\rm peak}$, motivated by the observation that the spectrum of $\phi$ is strongly peaked around $k_{\rm peak}$.
 \item We neglect the expansion of the universe because the timescales considered below are much shorter than $\Hubble^{-1}$. It is nevertheless possible that the expansion moves modes out of resonance if the resonance bands are very narrow. We will see in the lattice simulations presented in section~\ref{sec:num_analysis_lattice} whether the resonance is also effective when the expansion is taken into account.
\end{enumerate}
These assumptions are based on the main qualitative features of our system, and they allow us to derive simple equations that can be analysed using Floquet theory. We therefore believe that they provide a useful starting point for discussing how we can have parametric resonance even with an  inhomogeneous inflaton background field.

\subsection{Equations of motion}

The equation of motion for $\chi(\vec{x},t)$ is
\begin{align}
 \ddot{\chi} \, - \, \vec{\nabla}^2\chi \,+ \, 3H \dot{\chi} \, + \, \frac{\partial V}{\partial \chi} \, = \, 0.\label{eq:eomChi1}
\end{align}
As discussed above, we can neglect $H$ for the purposes of this section. For small perturbations $\delta\chi$ and $\delta \phi = (\phi - v)$, we can also expand eq.~\eqref{eq:eomChi1} to leading order in $\delta\chi$ and $\delta \phi$:
\begin{align}
 \delta\ddot{\chi} \, - \, \vec{\nabla}^2\delta\chi \, + \, \left[ \left.\frac{\partial^2 V}{\partial \chi^2}\right|_{\rm min} + \left.\frac{\partial^3 V}{\partial \chi^2 \partial \phi}\right|_{\rm min} \delta\phi \right] \delta\chi \, = \, 0,
\end{align}
where the derivatives of $V$ are evaluated at the global minimum $\phi = v$, $\chi = 0$:
\begin{align}
 \delta\ddot{\chi}(\vec{x},t) \, - \, \vec{\nabla}^2\delta\chi(\vec{x},t) \, + \, \left[ \lambda^2 v^4 + 4\lambda^2 v^3 \,\delta \phi(\vec{x},t) \right] \delta\chi(\vec{x},t) \, = \, 0.
\end{align}
We can Fourier transform this equation by multiplying with $e^{-i\vec{k}\vec{x}}$ and integrating over $d\vec{x}$:
\begin{align}
 \ddot{\chi}_{\vec{k}} \, + \, \left( k^2 + \lambda^2 v^4 \right) \chi_{\vec{k}} + 4\lambda^2 v^3 \!\int\! d\vec{x} ~ e^{-i\vec{k}\vec{x}} \delta \phi(\vec{x},t)\delta\chi(\vec{x},t) \, = \, 0,\label{eq:eomChi2}
\end{align}
where $\chi_{\vec{k}}(t)$ is the Fourier transform of $\delta \chi(\vec{x},t)$. To evaluate the last term, we approximate $\delta\phi$ as a standing wave with wavenumber $\lvert \vec{k_{\rm p}} \rvert = k_{\rm peak}$:
\begin{align}
 \delta \phi(\vec{x},t) \, = \, \delta\phi_0 \, \cos\left( \vec{k}_{\rm p} \vec{x} \right)\cos\left( \omega_\phi t \right), \quad~~~ \text{with }\, \omega_\phi \, = \, \sqrt{ k_{\rm peak}^2 + m_\phi^2 }. \label{eq:standingWave}
\end{align}
Note that eq.~\eqref{eq:standingWave} is a solution to the linearised equation of motion for $\delta\phi(\vec{x},t)$, so the approximation as a standing wave is consistent given our assumption that $\delta \phi \ll v$. Assuming this form of $\delta \phi$, eq.~\eqref{eq:eomChi2} can be simplified to
\begin{align}
 \ddot{\chi}_{\vec{k}} \, + \, \left( k^2 + \lambda^2 v^4 \right) \chi_{\vec{k}} + 2\lambda^2 v^3 \delta\phi_0 \cos(\omega_\phi t) \left( \chi_{\vec{k}+\vec{k}_{\rm p}} + \chi_{\vec{k}-\vec{k}_{\rm p}} \right) \, = \, 0.\label{eq:eomChi3}
\end{align}
This is similar to the usual Mathieu equation for parametric resonance, with the important difference that the time-dependent term couples modes with different $\vec{k}$.\footnote{For another generalisation of the Mathieu equation with inhomogeneous background fields in the context of colliding domain walls, see \cite{Braden:2014cra}.} For each $\vec{k}$, eq.~\eqref{eq:eomChi3} actually describes a ladder of differential equations involving all $\chi_{\vec{k}+n\cdot \vec{k}_{\rm p}}$ for every integer number $n$.

We cannot solve these infinitely many equations simultaneously, so we have to apply a cutoff, setting $\chi_{\vec{k}} = 0$ for all $\lvert \vec{k} \rvert > k_{\rm cutoff}$ or equivalently for all $\lvert n \rvert > N$.\footnote{Applying a cutoff is consistent with our lattice results, where we see that the spectrum of $\chi$ is peaked at values $k \lesssim k_{\rm peak}$, so we do not expect Fourier modes $\chi_k$ with $k \gg k_{\rm peak}$ to be involved in preheating. We have also checked that if we solve eq.~\eqref{eq:eomChi3} numerically using such a cutoff, the eigenfunctions of eq.~\eqref{eq:eomChi3} are peaked around one or two neighbouring values of $k$ and quickly fall off for larger $k$. In addition, we reproduced fig.~\ref{fig:floquetChi} with lower resolution for $k_{\rm cutoff} = 6k_{\rm peak}$ and found that the result does not change significantly.}

\subsection{Equations of motion in matrix form}

Imposing such a cutoff, eq.~\eqref{eq:eomChi3} can be written in matrix form. We use the short notation
\begin{align}
 \chi_{\low n} := \chi_{\low \vec{k}_n},  \quad\quad \vec{k}_n := \vec{k}_0 + n \cdot \vec{k}_{\rm p},
\end{align}
for any arbitrary momentum vector $\vec{k}_0$, and
\begin{align}
 f(t) \, := \, 2\lambda^2 v^3 \delta\phi_0 \cos(\omega_\phi t).
\end{align}
In this notation, eq.~\eqref{eq:eomChi3} becomes
\begin{align}
\begin{pmatrix}
 \ddot{\chi}_{\low N} \\
 \ddot{\chi}_{\low N-1} \\
 ... \\
 \ddot{\chi}_{\low -(N-1)} \\
 \ddot{\chi}_{\!\low -\!N}
 \end{pmatrix}
 \, &= \, 
\mathcal{F}
\begin{pmatrix}
 \chi_{\Low N} \\
 \chi_{\Low N-1} \\
 ... \\
 \chi_{\Low -(N-1)} \\
 \chi_{\!\Low -\!N}
\end{pmatrix},\label{eq:eomChi4}
\end{align}
with the matrix
\begin{align}
 \mathcal{F}(t) = -\lambda^2 v^4 \mathbb{1} - \begin{pmatrix}
 k_{N}^2 & f(t) & 0 & 0 & ...& 0\\
 f(t) & k_{N-1}^2 & f(t) & 0 & ...& 0\\
 0 & f(t) & k_{N-2}^2 & f(t) & ...& 0\\
 0 & 0 & f(t) & k_{N-3}^2 &... & 0\\
 ... & ... & ... & ... & ... & f(t) \\
 0 & 0 & 0 & 0 & f(t)  &k_{-N}^2\\
\end{pmatrix}.
\end{align}

As a last step, we transform eq.~\eqref{eq:eomChi4} into a first order differential equation, introducing extra variables
\begin{align}
 \pi_n \, := \, \dot{\chi}_n.
\end{align}
Then, for the vector
\begin{align}
y(t) := ( \chi_{\Low N}, \chi_{\Low N-1}, ..., \chi_{\Low -N}, \pi_{\Low N}, \pi_{\Low N-1}, ..., \pi_{\Low -N} )^T,
\end{align}
eq.~\eqref{eq:eomChi4} can be written as
\begin{align}
 \dot{y}(t) \, = \, U(t) y(t),\label{eq:eomChi5}
\end{align}
with
\begin{align}
 U(t) \, &= \,
 \begin{pmatrix}
 \mathbb{0} & \mathbb{1} \\
 \mathcal{F}(t) & \mathbb{0}
 \end{pmatrix}.\label{eq:U}
\end{align}

\subsection{Floquet analysis}

Eq.~\eqref{eq:eomChi5} is formally equivalent to equations of parametric resonance in a multi-field system, with different modes $\chi_{\vec{k}_0 + n \cdot \vec{k}_{\rm p}}$ in place of different fields $\varphi_n$. We can thus perform a Floquet analysis analogously to multi-field parametric resonance \cite{Amin:2014eta}.\footnote{Fundamentally, Floquet theory can be applied because the explicit time-dependence in $U(t)$ is periodic.}

The Floquet theorem implies that we can find the growing solutions of eq.~\eqref{eq:eomChi5} using the differential equation for the square matrix $\mathcal{O}(t)$:
\begin{align}
 \dot{\mathcal{O}}(t) \, = \, U(t) \mathcal{O}(t),\label{eq:O}
\end{align}
with $U(t)$ from eq.~\eqref{eq:U}. The Floquet exponents can be determined using the following algorithm \cite{Amin:2014eta}:
\begin{enumerate}
 \item Solve eq.~\eqref{eq:O} with the initial value $\mathcal{O}(0) = \mathbb{1}$ up to time $T = 2\pi/\omega_\phi$.
 \item Find the eigenvalues $\sigma$ of $\mathcal{O}(T)$.
 \item The Floquet exponents are $\mu = \frac{1}{T}\log \sigma$. Exponential growth happens for $\operatorname{Re}(\mu) = \frac{1}{T}\log\lvert \sigma \rvert > 0$. Taking into account Hubble damping, effective growth is possible for $\operatorname{Re}(\mu) \gg \Hubble$.
\end{enumerate}

\begin{figure}[tbp]
\hspace{-15pt}
$\begin{array}{ccc}
\includegraphics[width=0.51\textwidth]{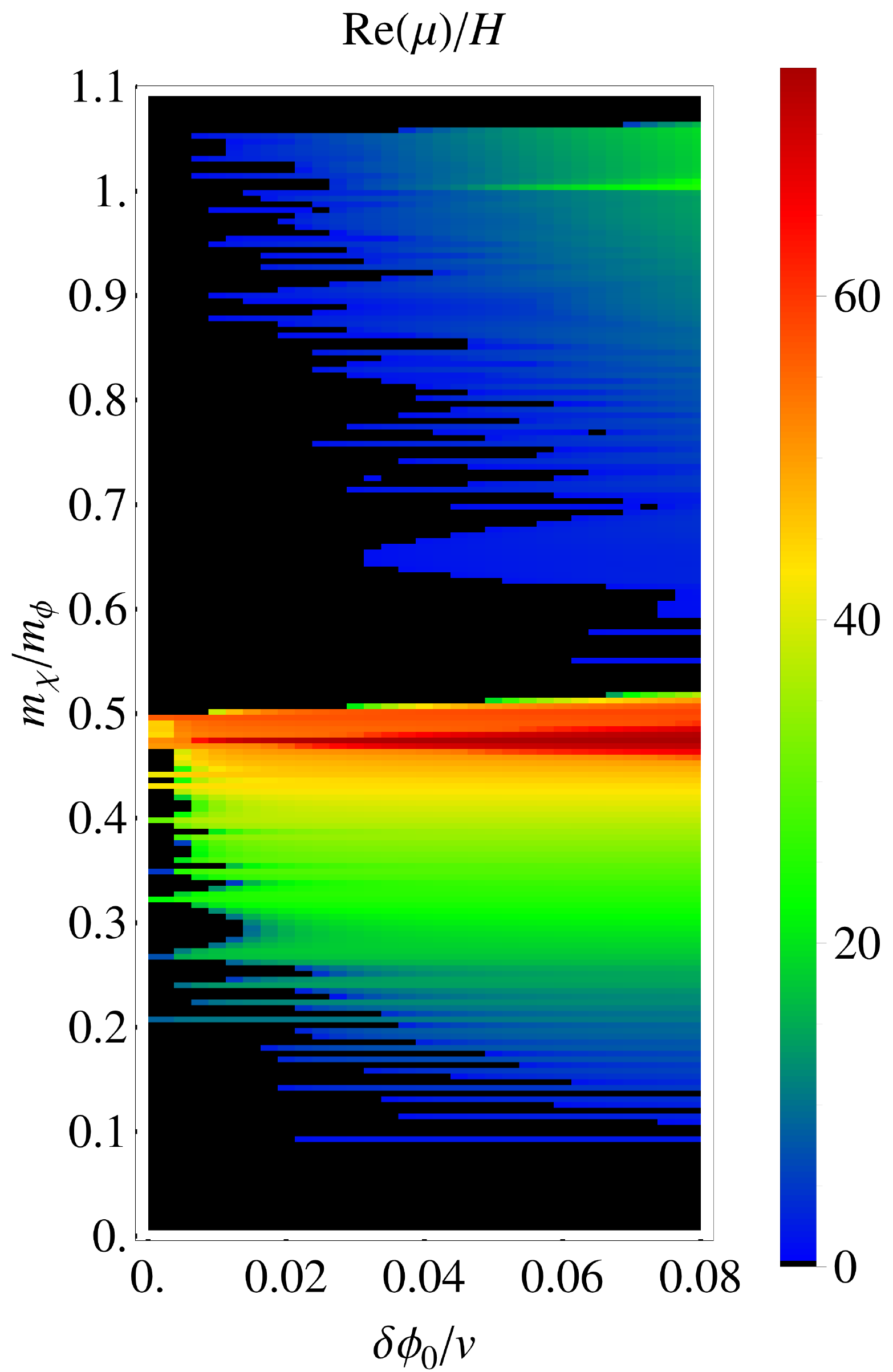} &
\includegraphics[width=0.515\textwidth]{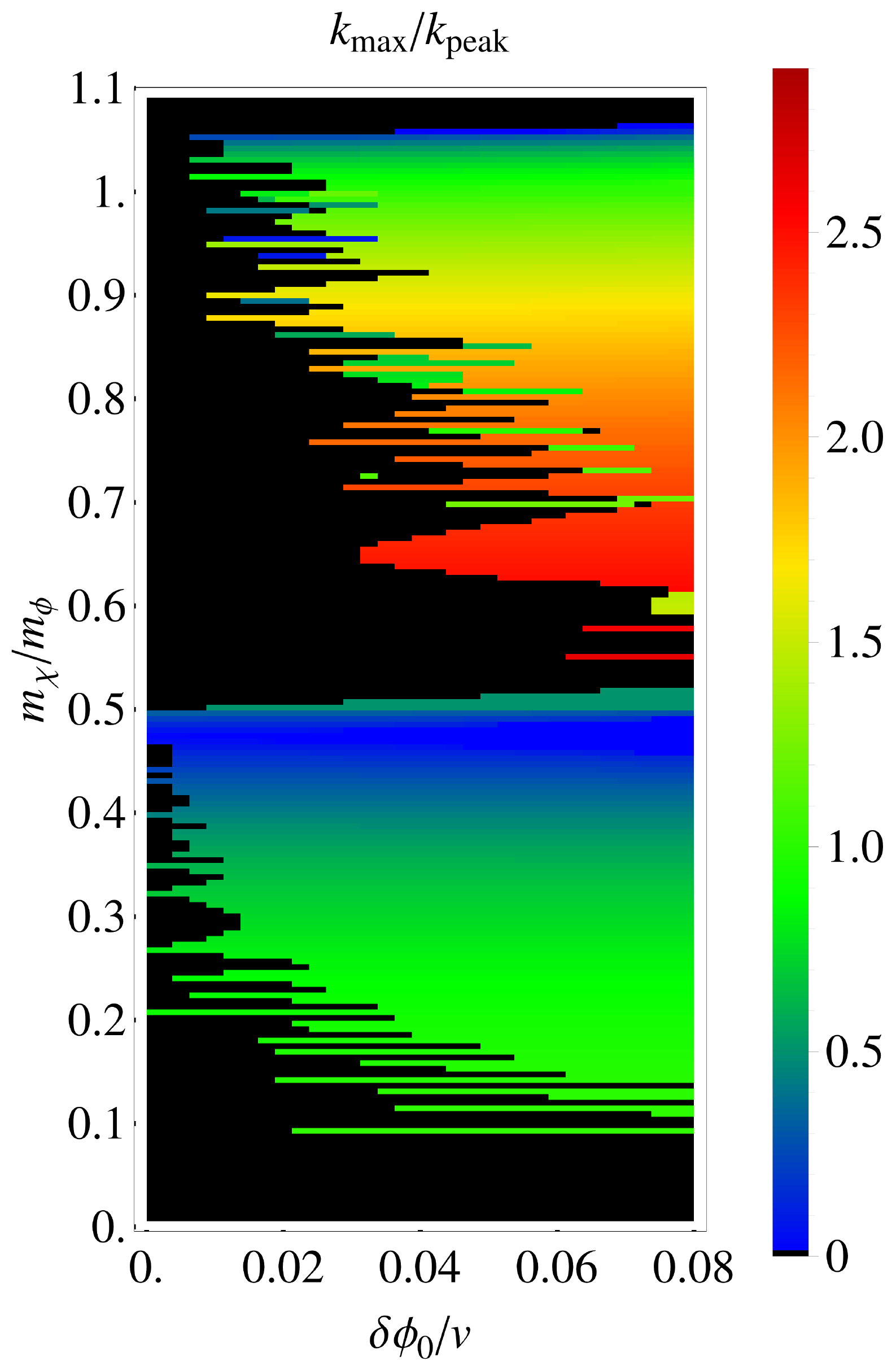}
\end{array}$
  \caption{Results of Floquet analysis for $v=10^{-2} \mpl$, using a cutoff $k_{\rm cutoff} = 3.5k_{\rm peak}$. The left plot shows the largest Floquet exponent for a given mass ratio $m_\chi/m_\phi$ and inflaton perturbation amplitude $\delta\phi_0$, indicating how strongly the fastest-growing $\chi$ perturbations are growing. Black regions indicate no growth (only oscillatory solutions), while in the coloured regions $\chi$ perturbations grow exponentially. The right plot shows the wavenumber $k_{\rm max}$ around which the fastest-growing linear combination of $\chi_k$ is peaked. The eigenvector generally also includes noticeable contributions with $(k_{\rm max} \pm k_{\rm peak})$, but the contributions to the eigenvector quickly fall off with larger $k$, justifying the use of a cutoff for solving the coupled eqs.~\eqref{eq:eomChi3}. Further faint bands can be found at $m_{\chi}/m_{\phi} \simeq 3/2$ and $m_{\chi}/m_{\phi} \simeq 2$, but the Floquet exponents are much smaller for those bands.\newline
  The presence of thin resonance bands is an artefact of our finite resolution in $k$-space, as we have only scanned over 41 different $k_0$, and the value of $k_0$ determines the position of the individual resonance bands. Including more $k_0$, the individual resonance bands combine into continuous broad bands.}
  \label{fig:floquetChi}
\end{figure}

The eigenvalues $\sigma$ describe how much the corresponding eigenstates grow during each oscillation of $\delta \phi$. Fig.~\ref{fig:floquetChi} shows the numerical results for the eigenvalues and eigenstates assuming $v=10^{-2} \mpl$ and different $m_\chi/m_\phi$ and $\delta\phi_0$. The left side of the plot shows the maximum Floquet exponent (for the fastest growing eigenfunction), and the right-hand side shows at which value of $k$ the corresponding eigenfunction is peaked.

To generate the plots, we scanned over 41 different $\vec{k}_0$ with $\lvert \vec{k}_0 \rvert \in [ 0, k_{\rm peak}/2 ]$ and $\vec{k}_0 \parallel \vec{k}_{\rm p}$.\footnote{We have also performed the Floquet analysis with angles of $30^\circ$ and $90^\circ$ between $\vec{k}_0$ and $\vec{k}_{\rm p}$, with $\lvert \vec{k}_0 \rvert \leq 1.5 k_{\rm peak}$. Such non-parallel $\vec{k}_0$ lead to a very similar band structure, just with slightly broader bands.} As eq.~\eqref{eq:eomChi3} couples each $\vec{k}_0$ to all other $\vec{k}_0 + n \cdot \vec{k}_{\rm p}$, and because the equations are symmetrical under $\vec{k}_0 \rightarrow -\vec{k}_0$, this is equivalent to scanning over all $\lvert \vec{k}_0 \rvert \leq k_{\rm cutoff}$ for $\vec{k}_0 \parallel \vec{k}_{\rm p}$.

For each of the $k_0$, we find a series of thin resonance bands for specific $m_\chi/m_\phi$. Plotting all of these thin resonance bands in one plot (plotting the maximum eigenvalue), we find that they combine into very broad bands with strongest amplification close to $m_{\chi}/m_{\phi} \in \{0.5, 1, 1.5, 2\}$. These broad bands extend towards smaller $m_\chi/m_\phi$, with the exponentially growing modes for smaller $m_\chi/m_\phi$ having increasingly larger $k$, and the growth getting weaker the further we go away from half-integer values of $m_{\chi}/m_{\phi}$. Also, the strongest amplification happens for $m_{\chi}/m_{\phi} \lesssim 0.5$. The band at $m_{\chi}/m_{\phi} \simeq 1$ already has much smaller Floquet exponents, and the bands at $m_{\chi}/m_{\phi} \simeq 1.5$ and $m_{\chi}/m_{\phi} \simeq 2$ only exhibit extremely weak amplification.

Note that our plane-wave approximation for $\delta\phi$ requires $V(\delta\phi) \simeq \frac12 m_\phi^2\delta\phi^2$. For this reason, the results of our Floquet analysis will be most trustworthy for small $\delta\phi_0$, i.e.\ on the left side of the plots in fig.~\ref{fig:floquetChi}. The lower band at $m_\chi/m_\phi \lesssim 0.5$ extends to small $\delta\phi_0$, with $\operatorname{Re}(\mu)/\Hubble > 20$ even for $\delta\phi_0/v < 0.01$ where the potential is very close to harmonic. However, the band at $m_\chi/m_\phi \simeq 1$ becomes very faint already for $\delta\phi_0/v \sim 0.05$ where our actual potential~\eqref{eq:potential} for $\delta\phi$ is still significantly skewed. For this reason, it is not clear whether the conditions for our Floquet analysis are satisfied sufficiently well for this upper band given the anharmonic hilltop potential of eq.~\eqref{eq:potential}.

Our Floquet analysis demonstrates that, in principle, exponential growth of $\chi$ due to parametric resonance might happen even for an inhomogeneous background field $\phi$ with a sharp peak around some $k_{\rm peak}$, and that such growth sensitively depends on the coupling $\lambda$ or equivalently the mass ratio $m_\chi/m_\phi$. However, since our analysis is based on a harmonic approximation of the inflaton potential and on simpler initial conditions for $\phi$ compared to the hilltop model of eq.~\eqref{eq:potential}, we need to resort to lattice simulations to assess whether or not this mechanism is effective in our hilltop model.

\section{Results of lattice simulations}
\label{sec:num_analysis_lattice}
In this section we present results of numerical lattice simulations. For $v=10^{-2}\,m_{\rm{Pl}}$, we simulated the evolution of the model~\eqref{eq:potential} during the first few $e$-folds of preheating. To this end, we used a modified version of LATTICEEASY \cite{Felder:2000hq}. The program solves the following system of non-linear equations on a discrete spacetime lattice:
\begin{align}
\ddot{\phi}(t,\overline{x})\, + \, 3H\dot{\phi}(t,\overline{x})\, &- \,\frac{1}{a^2}\nabla_{\overline{x}}^2\phi(t,\overline{x})\, + \,\frac{\partial V}{\partial\phi} \, = \, 0\,,\\
\ddot{\chi}(t,\overline{x})\, + \, 3H\dot{\chi}(t,\overline{x})\, &- \,\frac{1}{a^2}\nabla_{\overline{x}}^2\chi(t,\overline{x})\, + \,\frac{\partial V}{\partial\chi} \, = \, 0\,, \\
H^2\, = \,\frac{1}{3m^2_{\rm{Pl}}}\bigg\langle V\,  + \,\frac{1}{2}\left(\dot{\phi}^2\,  + \,\dot{\chi}^2\right)& + \frac{1}{2a^2}\left(\left|\nabla_{\overline{x}}\phi\right|^2  \, + \, \left|\nabla_{\overline{x}}\chi\right|^2\right)\bigg\rangle\,,
\label{eq:EOMs_phi}
\end{align}
where the angle brackets denote averages over the whole grid and $\overline{x}$ are comoving coordinates. 

Most of our simulations were performed in $2+1$ dimensions with $N=2048$ lattice points per dimension. For the plot in fig.~\ref{fig:ratios}, however, we used $N=1024$ due to limited computing power. We checked that reducing the number of points does not significantly change the results.

\subsection{Initial conditions of the lattice simulations}
\label{sec:lattice_initial_conditions}

The initial conditions for the fields as well as the configuration of the lattice are summarized in table~\ref{tab:ic_and_parameters}. $\phi$ and $\chi$ are initialized at the end of inflation, such that the initial field fluctuations within the lattice are still well-described by vacuum fluctuations. The initial field derivative $\braket{\dot{\phi}}_{\rm i}$ is calculated from the homogeneous field equations. In this section we assume zero initial conditions for $\braket{\chi}$. The effects of non-zero initial conditions for $\braket{\chi}$ are briefly discussed in appendix~\ref{sec:nonzerochi}.

The Fourier components of the fields and their derivatives are initialized as
\begin{align}
f_k \, &= \, \frac{1}{a}\frac{\lvert f_k \rvert}{\sqrt{\alpha^2_++\alpha^2_-}}\left(\alpha_+\,e^{i2\pi\theta_++ikt}+\alpha_-\,e^{i2\pi\theta_--ikt} \right)\,,\\
\dot{f}_k \, &= \, \frac{ik}{a}\frac{\lvert f_k \rvert}{\sqrt{\alpha^2_++\alpha^2_-}}\left(\alpha_+\,e^{i2\pi\theta_++ikt}-\alpha_-\,e^{i2\pi\theta_--ikt} \right)-Hf_k\,,
\end{align}
where $\alpha_\pm$ and $\theta_\pm$ are real random numbers uniformly distributed between 0 and 1.
Here, the initialization differs from the one in the original LATTICEEASY code which uses $\alpha_\pm=1$ instead. The norm of the Fourier components is initialized as a Rayleigh distributed random variable, i.e.
\begin{align}
P(|f_k|)\, = \, \frac{2|f_k|}{\langle f^2_k\rangle_{\rm{vac}}}\,\exp{\left[-\frac{|f_k|^2}{\langle f^2_k\rangle_{\rm{vac}}}\right]}\,.
\label{eq:rayleigh}
\end{align}

The program uses rescaled variables to simplify the calculations. For our model, we chose the following rescalings for the fields $f$, the coordinates $\overline{x}$ and the time step $dt$:
\begin{align}
f_{\rm pr} \, = \, a\,f\,;\quad \overline{x}_{\rm pr}\, = \, 300H_{\rm i}\,\overline{x}\,;\quad dt_{\rm pr} \, = \, \frac{300H_{\rm i}}{a}\,dt.
\end{align}

\begin{table}[ht]
\begin{center}
\begin{tabular}{|c|c|c|c|c|c|}
  \hline
  \multicolumn{6}{|l|}{Initial conditions and model parameters} \\
  \hline
  \hline
  $v/m_{\rm Pl}$ & $\langle\phi\rangle_{\rm i}/v$ & $\langle\dot{\phi}\rangle_{\rm i}/v^2$ & $\langle\chi\rangle_{\rm i}$ & $\langle\dot{\chi}\rangle_{\rm i}$ & $H_{\rm i}/\mpl$\\
  \hline
 $10^{-2}$ & $8\times10^{-2}$ & $2.49\times10^{-9}$ & $0$ & $0$ & $1.87\times10^{-10}$\\
 \hline
  \hline
  \multicolumn{6}{|l|}{Lattice configuration} \\
  \hline
  \hline
 \multicolumn{2}{|c|}{\# of dimensions} & $N$ & $L\,H_{\rm i}$ &  $k_{\rm uv}$ & $k_{\rm ir}$\\
  \hline
 \multicolumn{2}{|c|}{$2+1$} & $2048$ & $0.21$ & $120\,k_{\rm peak}$ & $0.06\,k_{\rm peak}$\\
 \hline
 \end{tabular}
 \caption{Initial conditions and configuration of the lattice. The fields are initialized at the end of inflation and their derivatives are calculated from the field equations of motion, without assuming slow-roll. In this section we assume zero initial conditions for $\langle\chi\rangle$. The case of non-zero initial conditions for $\langle\chi\rangle$ is discussed in appendix~\ref{sec:nonzerochi}.}
  \label{tab:ic_and_parameters}
\end{center}
\end{table}

\subsection{Results for different $m_{\chi}/m_{\phi}$}
\label{sec:2048p}
In this section we present and discuss the results of lattice simulations for three selected values of $m_{\chi}/m_{\phi}$. We simulated the evolution during the non-linear stage of preheating for $m_{\chi}/m_{\phi} \approx 0.364$, $m_{\chi}/m_{\phi} = 0.5$ and $m_{\chi}/m_{\phi} = 1$, which  corresponds to setting  $\lambda = 1\times10^{-3}\,m^{-1}_{\rm Pl}$, $\lambda \approx 1.375\times10^{-3}\,m^{-1}_{\rm Pl}$ and $\lambda \approx 2.750\times10^{-3}\,m^{-1}_{\rm Pl}$, respectively. From our Floquet analysis in section~\ref{sec:floquet} we expect a strong enhancement for the first two mass ratios, while for $m_{\chi}/m_{\phi} = 1$ the amplification of $\chi$ fluctuations is expected to be much weaker if at all present (cf.\ fig.~\ref{fig:floquetChi}). The simulations were performed in $2+1$ dimensions, on a lattice with $2048$ points per dimension.

Below, particular attention is given to the evolution of the variances of the fields, since knowing their value at the end of preheating can be very useful to study the subsequent reheating. Particularly at late times, when the amplitudes of the fields are small enough such that the potential can be approximated by a quadratic potential and $k/a\ll m_f$ due to redshifting, the virial theorem states that $\braket{V(\delta f)} = \langle\dot{f}^2/2\rangle$, where $\delta f$ stands for the amplitude of the field $f$. The mean energy density is then closely related to the variance: $\langle\rho_f\rangle=m^2_f\langle\delta f^2\rangle$.

\subsection*{Variances}

\begin{figure}[tbp]
\centering
\subfigure{\includegraphics[width=7cm]{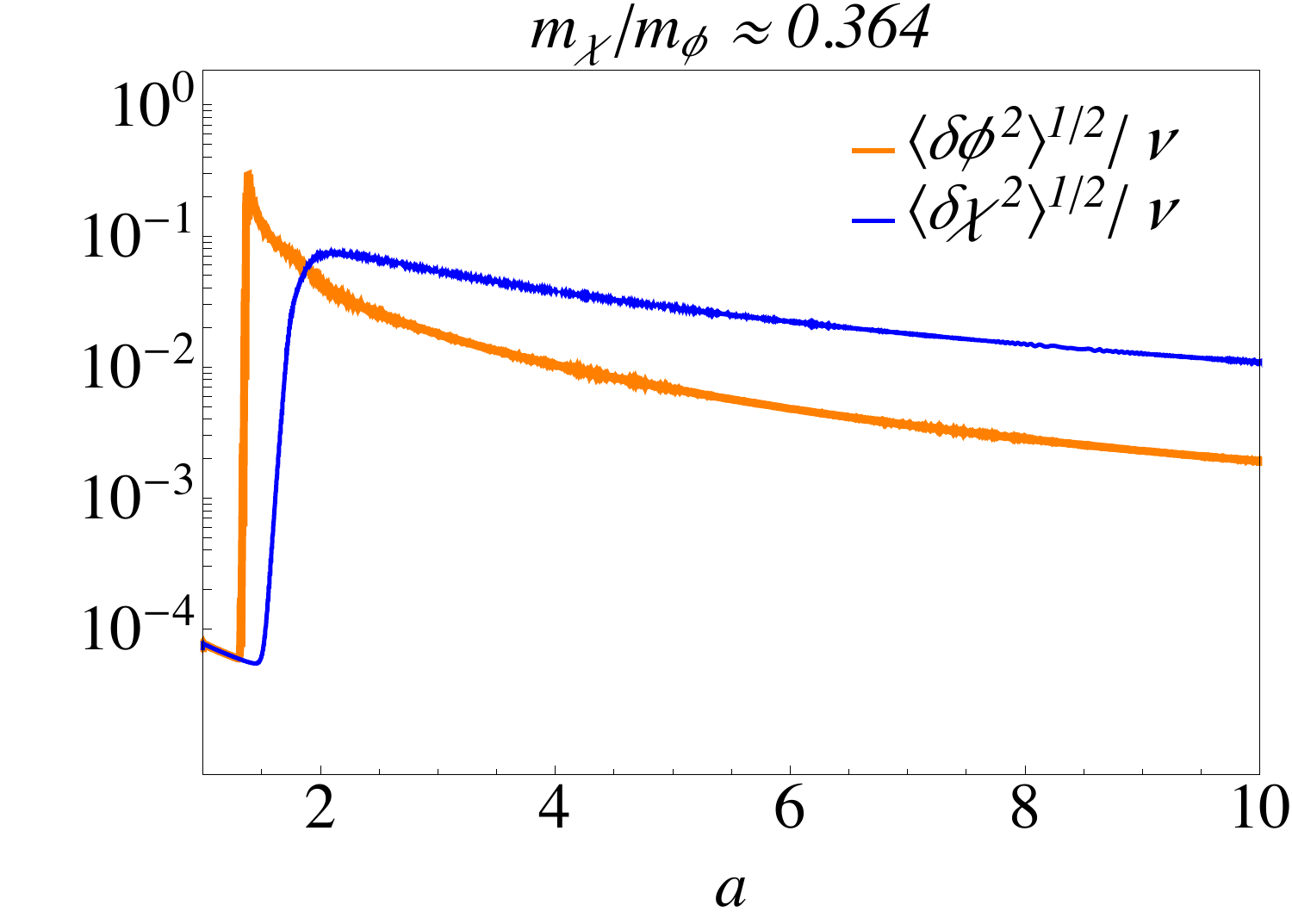}}
\hfill
\subfigure{\includegraphics[width=7cm]{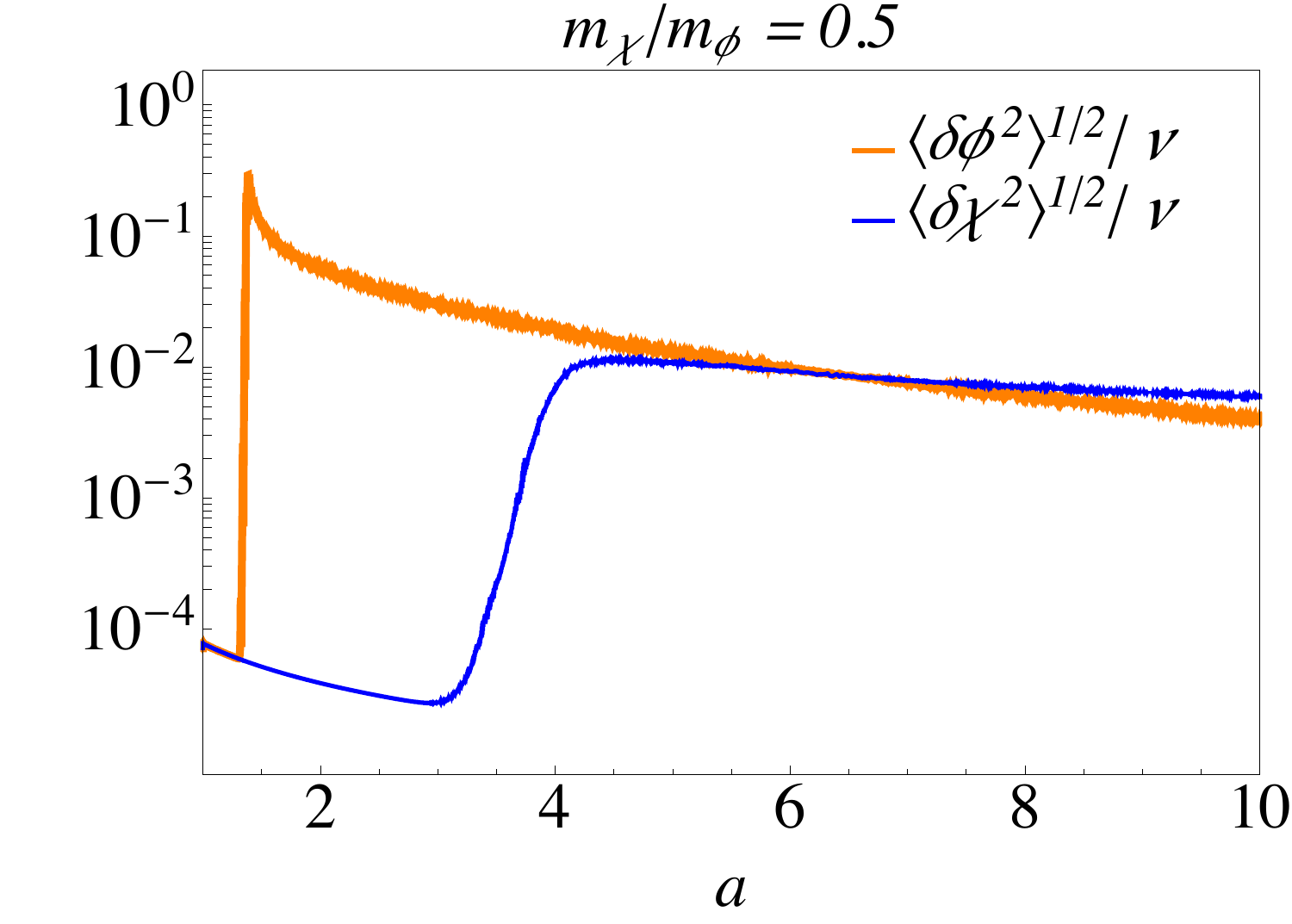}}
\hfill
\subfigure{\includegraphics[width=7cm]{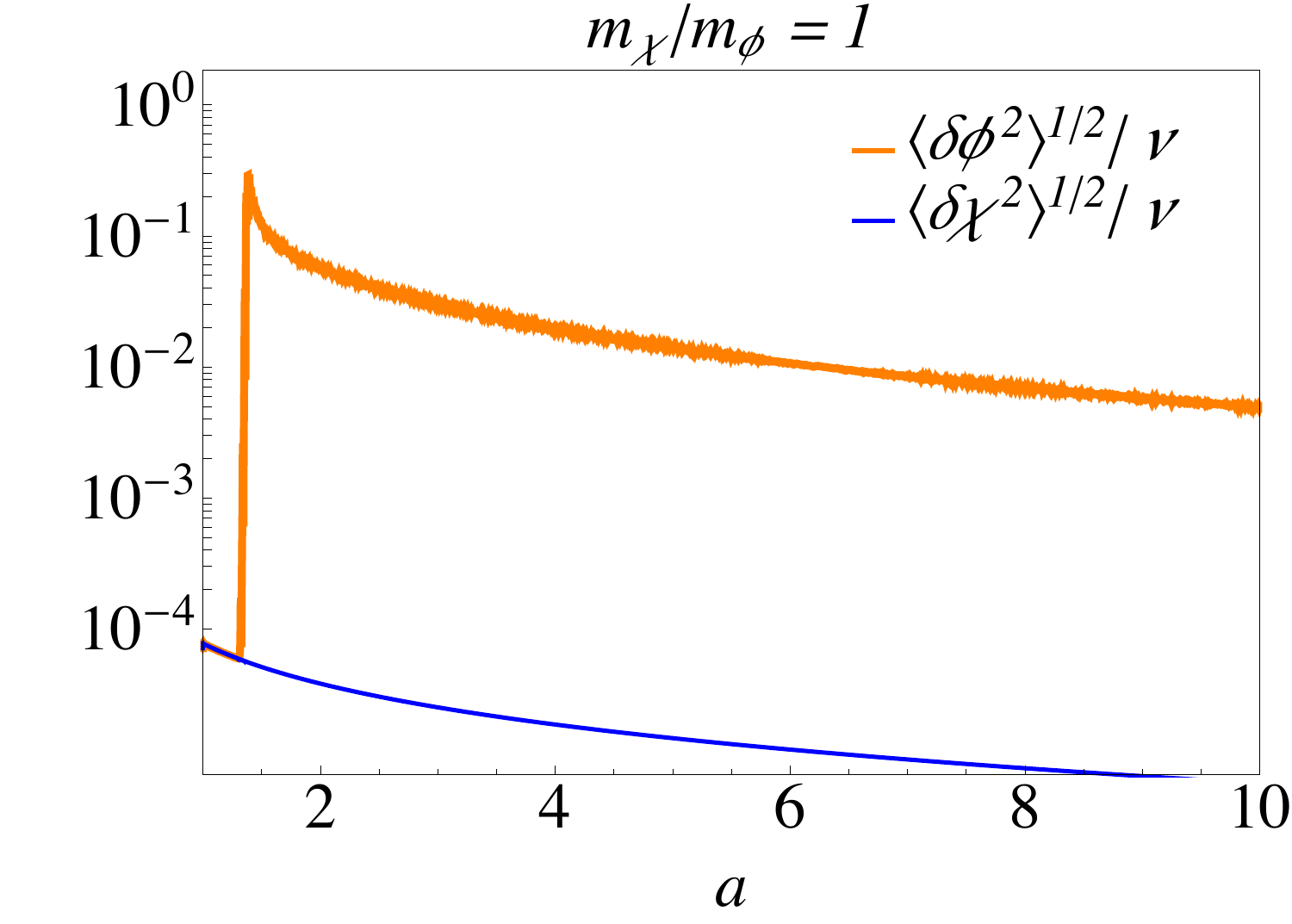}}
\hfill
\subfigure{\includegraphics[width=7cm]{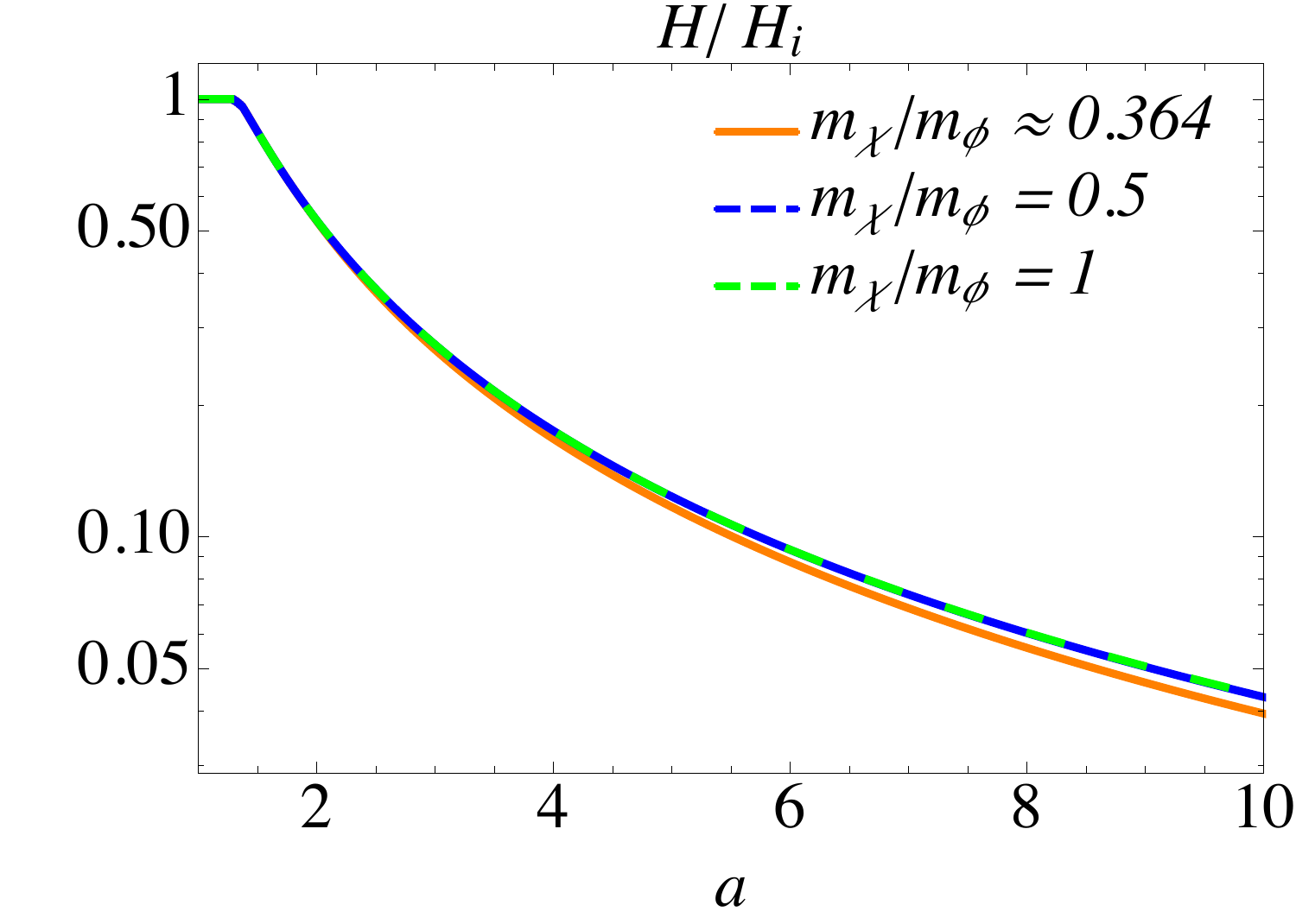}}
\caption{Results of lattice simulations for $v=10^{-2}m_{\rm Pl}$ and $m_{\chi}/m_{\phi}\approx0.364$, $m_{\chi}/m_{\phi}=0.5$ and $m_{\chi}/m_{\phi}=1$. The first three plots show the variances $\sqrt{\braket{ \delta\phi^2}}$ and $\sqrt{\braket{\delta\chi^2}}$ as functions of the scale factor $a$, and the lower right plot shows the evolution of the Hubble parameter $H(a)$.
In all three cases, $\sqrt{\braket{\delta\phi^2}}$ evolves qualitatively the same. The variance is initially amplified due to tachyonic oscillations which are terminated by non-linearities, after which the inflaton develops localized bubbles which oscillate between $\phi \sim -v$ and $\phi \sim +v$ (cf.\ \cite{Antusch:2015nla}). The transition to the non-linear dynamics corresponds to the ``spike" in the evolution of $\sqrt{\braket{\delta\phi^2}}$. On the other hand, the evolution of $\sqrt{\braket{\delta\chi^2}}$ significantly depends on the mass ratio $m_{\chi}/m_{\phi}$. Note that in those cases in which $\sqrt{\braket{\delta\chi^2}}$ is amplified, the growth happens after the tachyonic oscillations, when $\phi$ is dominated by inhomogeneities.
}
\label{fig:vars}
\end{figure}

Fig.~\ref{fig:vars} shows the evolution of $\sqrt{\langle \delta\phi^2 \rangle}/v$ and $\sqrt{\langle \delta\chi^2 \rangle}/v$ in terms of $a$, for $m_{\chi}/m_{\phi} \approx 0.364$ (left), $m_{\chi}/m_{\phi} = 0.5$ (middle) and $m_{\chi}/m_{\phi} = 1$ (right). For the first two mass ratios, $\sqrt{\langle \delta\chi^2 \rangle}$ experiences a phase of strong amplification, which happens after the tachyonic $\phi$ oscillations when the system is dominated by non-linearities. In both simulations $\sqrt{\langle \delta\chi^2 \rangle}$ grows larger than $\sqrt{\langle \delta\phi^2 \rangle}$. For $m_{\chi}/m_{\phi} \approx 0.364$, the growth happens around $a\approx1.4$ while for $m_{\chi}/m_{\phi} = 0.5$ it happens later, around $a\approx3$. The two cases not only differ with regard to the time at which the growth happens, but also the strength of the amplification is different. In the simulation with $m_{\chi}/m_{\phi} \approx 0.364$, $\sqrt{\langle \delta\chi^2 \rangle}$ is amplified more while $\sqrt{\langle \delta\phi^2 \rangle}$ has a smaller final value compared to the simulation with $m_{\chi}/m_{\phi} = 0.5$. This is explained by the stronger growth in $\sqrt{\langle \delta\chi^2 \rangle}$ which means that the inflaton transfers more energy to $\chi$.

Compared to the results of the Floquet analysis in section~\ref{sec:floquet}, the first two simulations shown in fig.~\ref{fig:vars} lie within the lower resonance band (cf.\ fig.~\ref{fig:floquetChi}), and the lattice results confirm the expectations from the semi-analytical estimates of section~\ref{sec:floquet}. Fig.~\ref{fig:floquetChi} also shows some indication of a weaker resonance at $m_{\chi}/m_{\phi}\sim1$, though only for larger inflaton amplitudes $\delta\phi/v$ for which the Floquet analysis is less reliable (see the discussion in section~\ref{sec:floquet} for details). In the more accurate lattice simulations, $\sqrt{\langle \delta\chi^2 \rangle}$ did not experience any growth for $m_\chi=m_\phi$ as one can see in fig.~\ref{fig:vars}. Also for other mass ratios around $m_{\chi}/m_{\phi}\sim1$ we did not observe any amplification of $\sqrt{\langle \delta\chi^2 \rangle}$.

\subsection*{Spectra}
\begin{figure}[tbp]
\centering
\subfigure{\includegraphics[width=3cm]{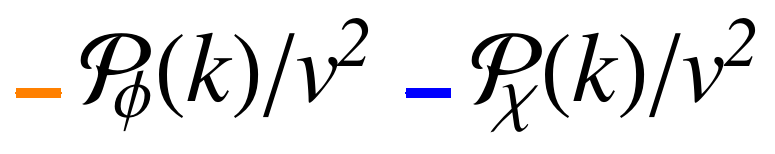}}\\
\subfigure{\includegraphics[width=5cm]{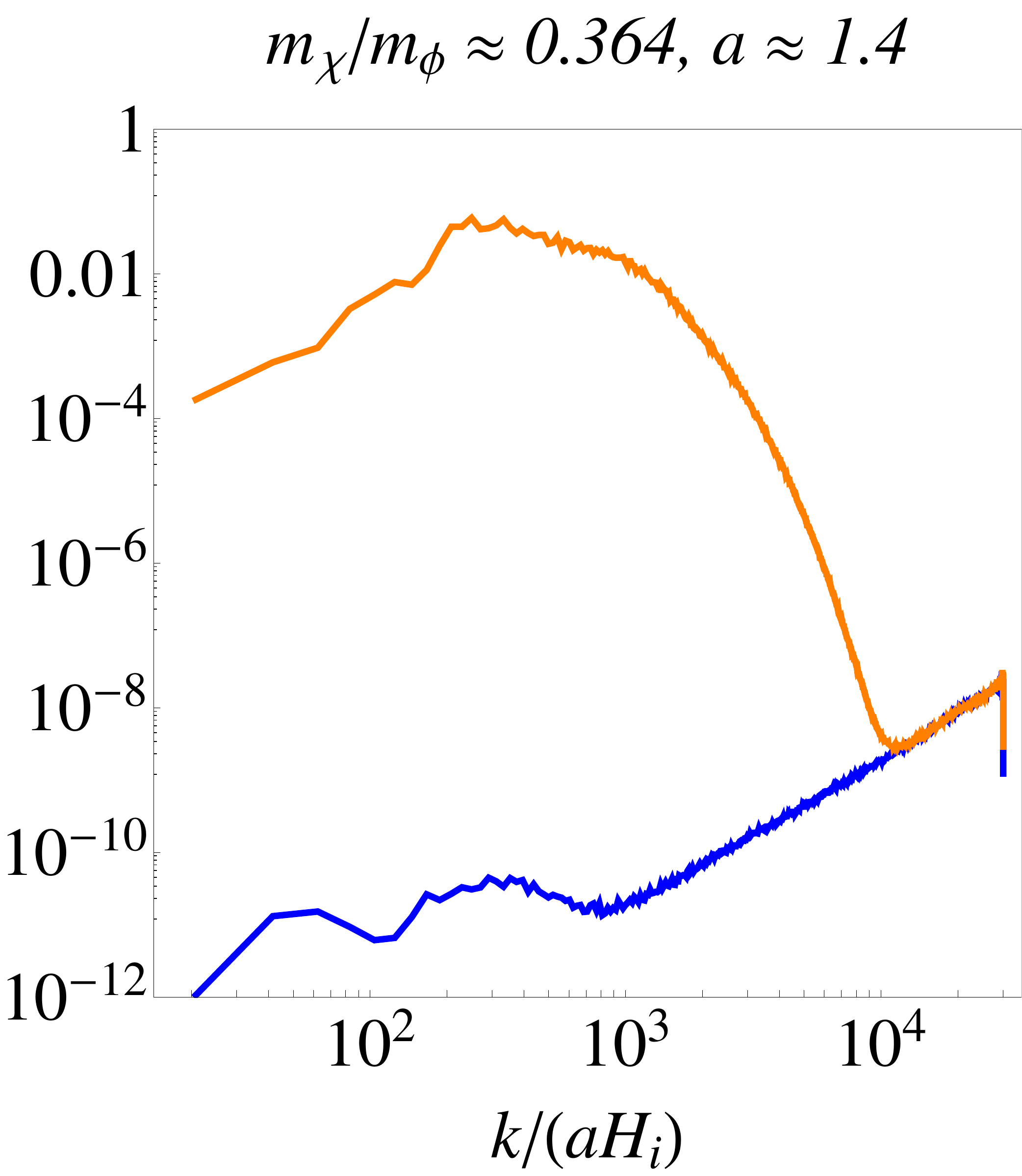}}
\hfill
\subfigure{\includegraphics[width=5cm]{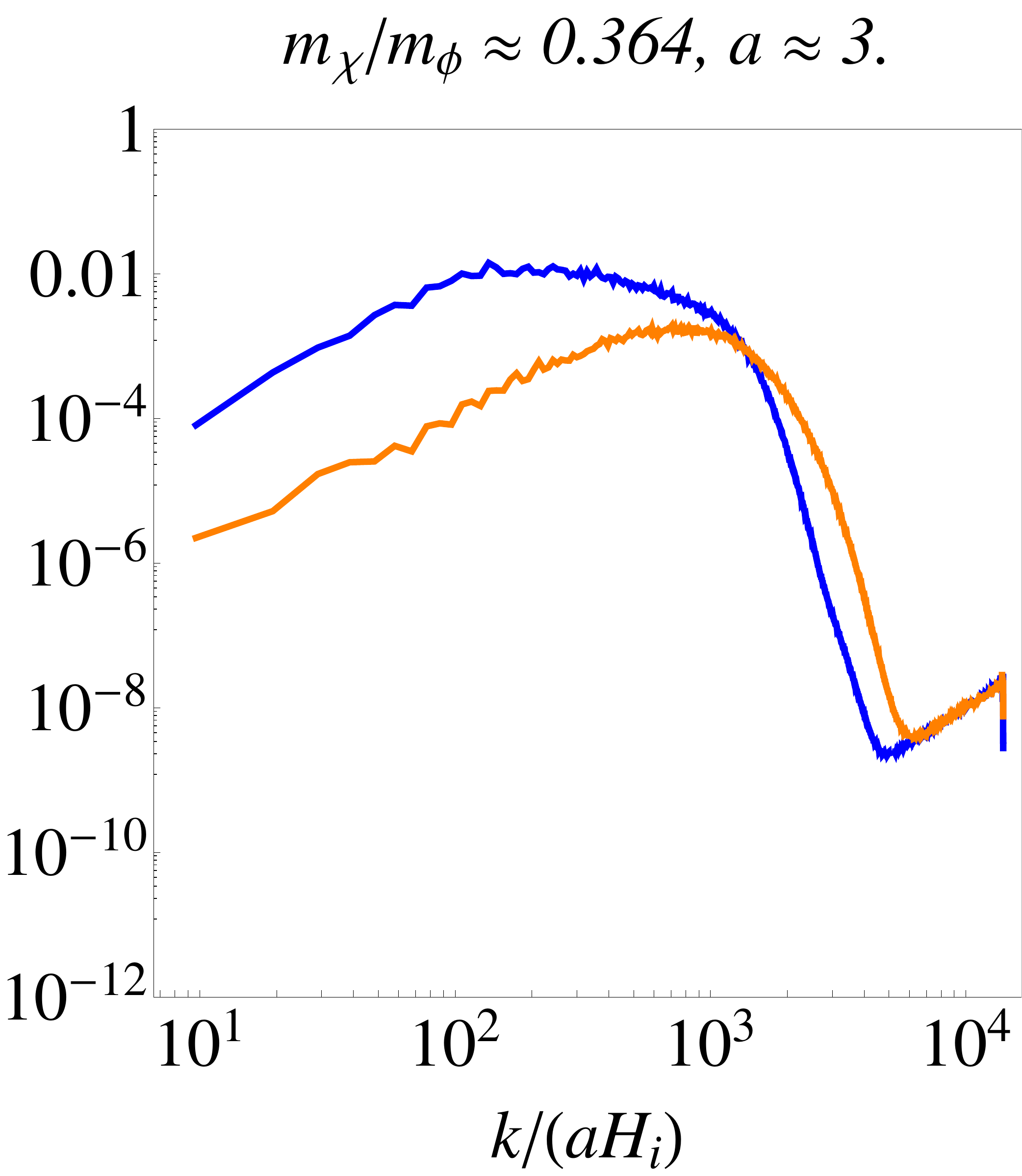}}
\hfill
\subfigure{\includegraphics[width=5cm]{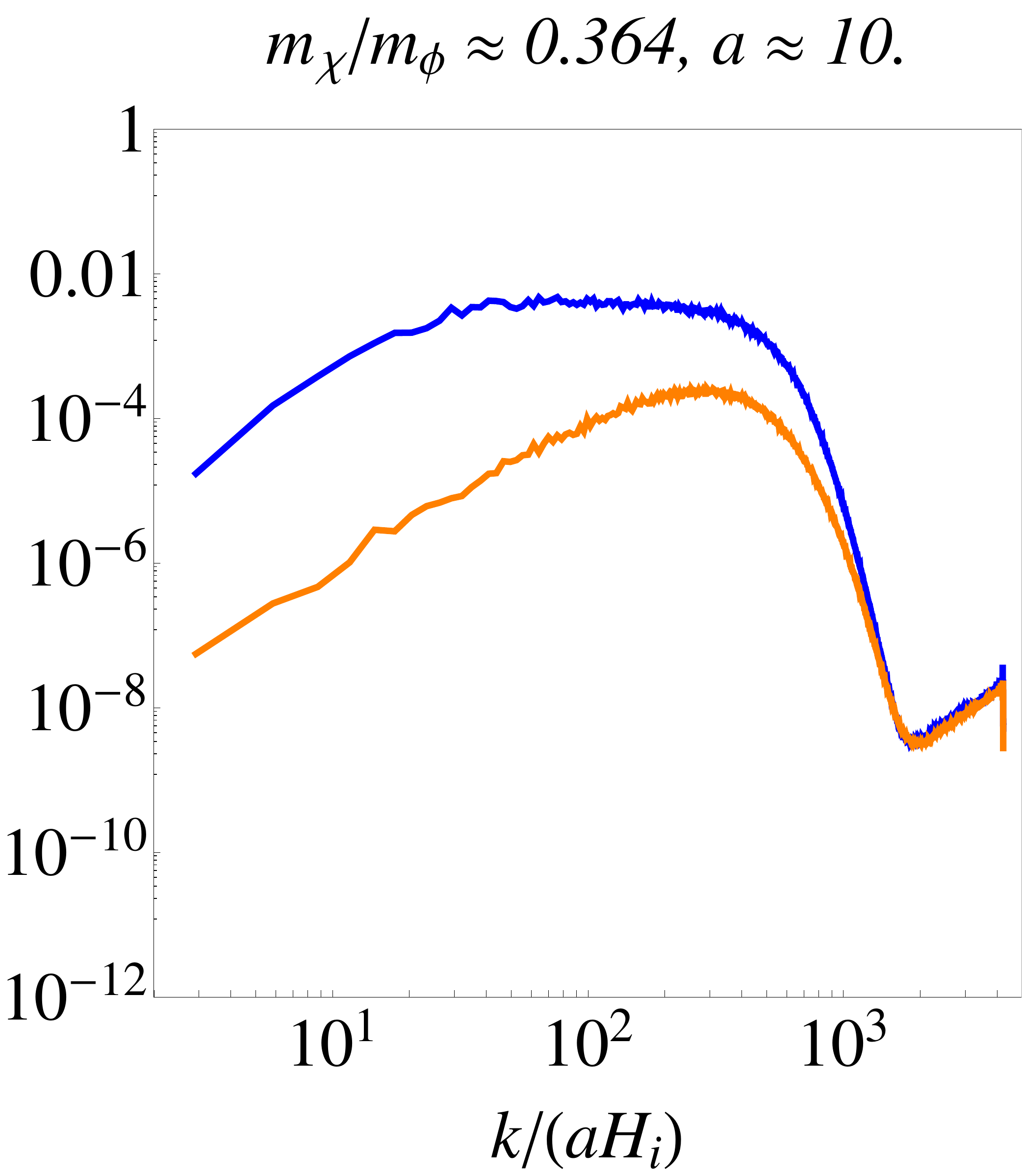}}
\hfill
\subfigure{\includegraphics[width=5cm]{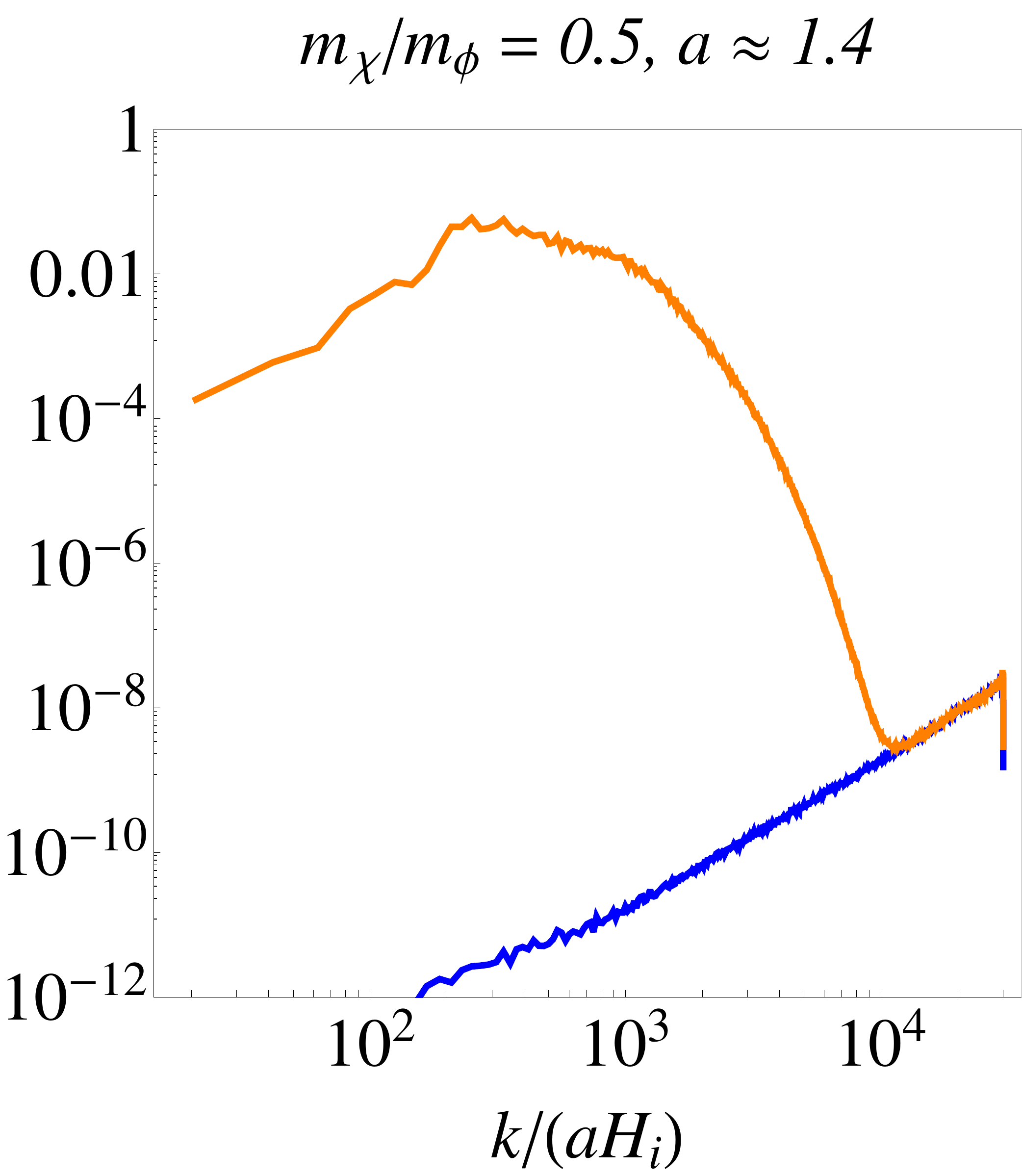}}
\hfill
\subfigure{\includegraphics[width=5cm]{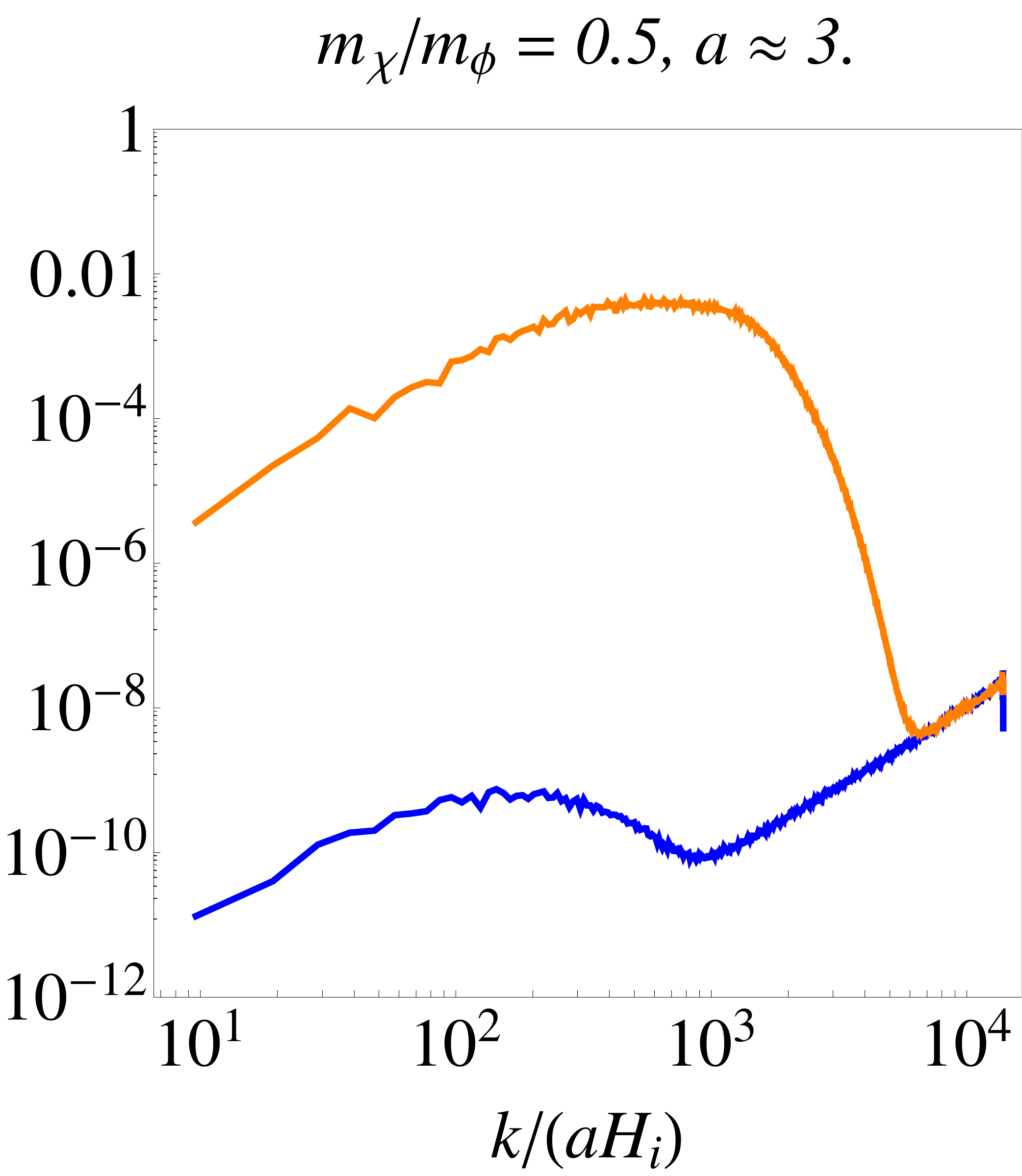}}
\hfill
\subfigure{\includegraphics[width=5cm]{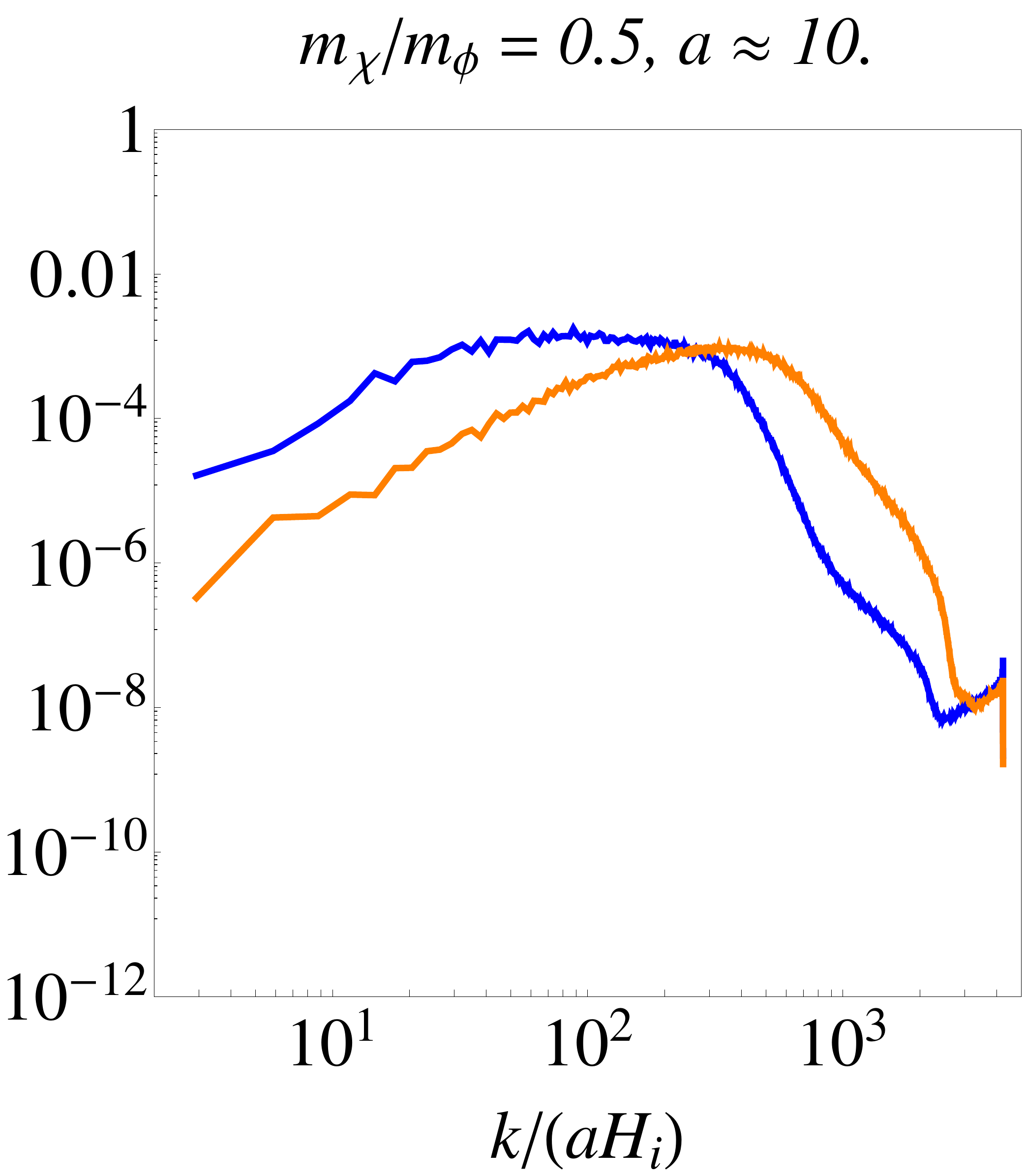}}
\hfill
\subfigure{\includegraphics[width=5cm]{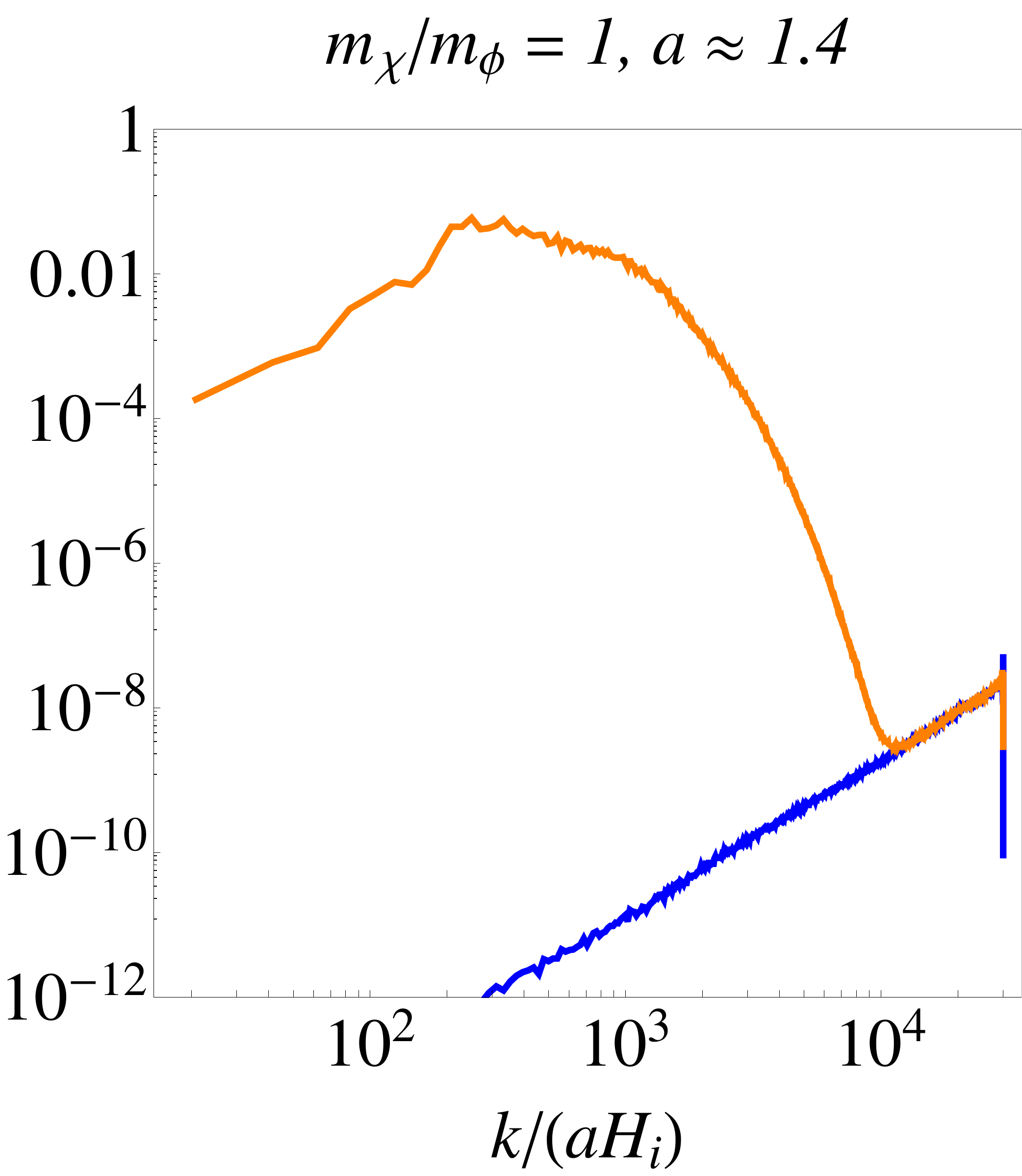}}
\hfill
\subfigure{\includegraphics[width=5cm]{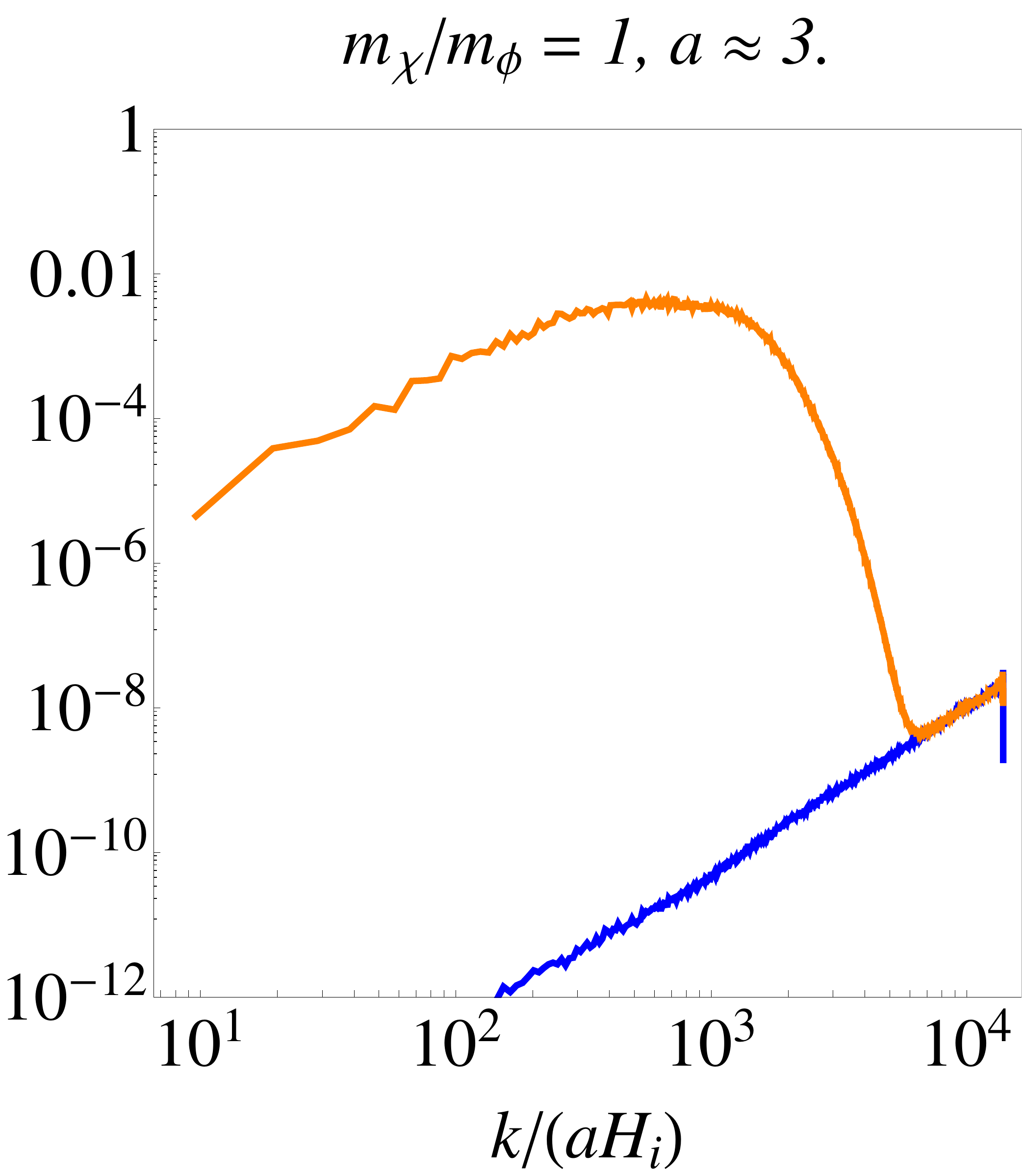}}
\hfill
\subfigure{\includegraphics[width=5cm]{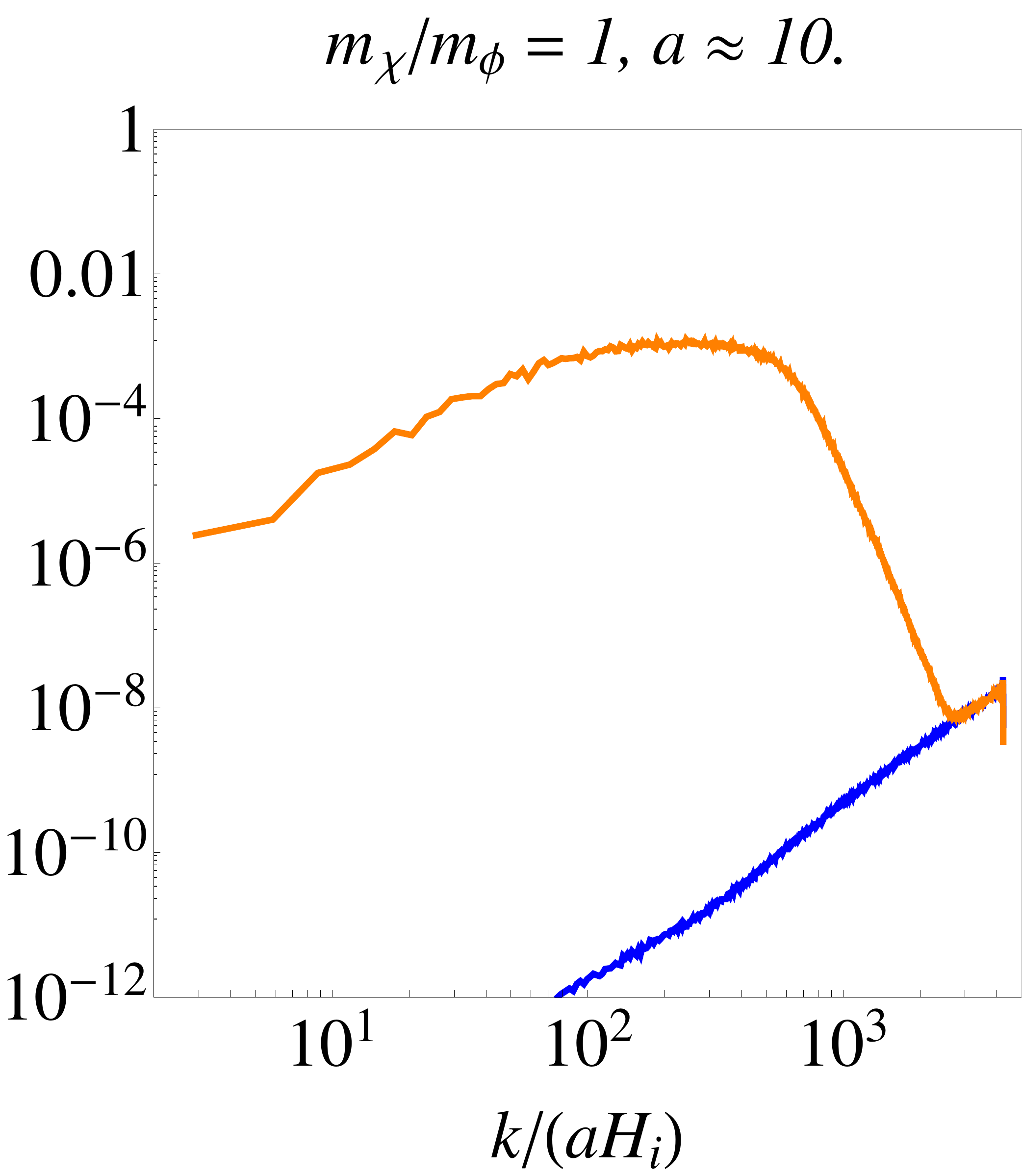}}
\caption{Results of lattice simulations for $v=10^{-2}m_{\rm Pl}$ and $m_{\chi}/m_{\phi}\approx0.364$ (upper three), $m_{\chi}/m_{\phi}=0.5$ (middle three) and $m_{\chi}/m_{\phi}=1$ (lower three). For each mass ratio, we show the spectra of fluctuations $\mathcal{P}_{\phi}(k)=\frac{k^3}{2\pi^2}|\phi_k|^2$ and $\mathcal{P}_{\chi}(k)=\frac{k^3}{2\pi^2}|\chi_k|^2$, evaluated at three different points in time, corresponding to (from left to right): $a\approx1.4$, $a\approx3$ and $a\approx10$. For all three mass ratios one can see the typical peak of the $\phi$ spectrum due to the tachyonic oscillations. On the other hand, the evolution of the $\chi$ perturbations depends on the value of $m_{\chi}/m_{\phi}$.}
\label{fig:spectra}
\end{figure}

The spectra $\mathcal{P}_{\phi}(k)=\frac{k^3}{2\pi^2}|\phi_k|^2$ and $\mathcal{P}_{\chi}(k)=\frac{k^3}{2\pi^2}|\chi_k|^2$ are illustrated in fig.~\ref{fig:spectra} for $m_{\chi}/m_{\phi} \approx 0.364$ (upper three), $m_{\chi}/m_{\phi} = 0.5$ (middle three) and $m_{\chi}/m_{\phi} = 1$ (lower three). For each value of $m_{\chi}/m_{\phi}$, we show three different plots corresponding to three different points in time: $a\approx1.4$, after non-linearities have terminated the tachyonic oscillations phase; $a\approx3$; and at the end of our simulation $a\approx10$.

For $m_{\chi}/m_{\phi}\approx0.364$, the amplification of $\chi$ perturbations starts very early. Shortly after the end of the tachyonic oscillations phase, around $a\approx1.4$, one can see a sharp peak in the spectrum of $\phi$ around the scale $k/(a\,H_{\rm i})\simeq300$. Furthermore, one can see that the $\chi$ perturbations already experienced an amplification compared to the vacuum spectrum (see e.g.\ the figures below). At $a\approx3$, both spectra show a sharp peak, with the peak in the $\chi$ spectrum about one order of magnitude larger, more extended and slightly shifted towards the infrared compared to the one in the $\phi$ spectrum. During the subsequent evolution until the end of our simulation at $a\approx10$, the peak of the $\chi$ spectrum remains more than one order of magnitude larger than the one of the $\phi$ spectrum. Moreover, the $\chi$ spectrum is broader towards smaller $k$ values compared to the spectrum of $\phi$ perturbations. 

For $m_{\chi}/m_{\phi} = 0.5$, in contrast to the simulation with $m_{\chi}/m_{\phi}\approx0.364$, the amplification of $\chi$ modes starts later, as one would expect from the variances in fig.~\ref{fig:vars}. At $a\approx3$, the amplification of $\chi$ fluctuations has already started and one can see a small peak developing around $k/(a\,H_{\rm i})\simeq100$ -- $200$. At the end of our simulation at $a\approx10$, the fields have both developed a relatively sharp peak and both peaks have the same order of magnitude. Again, the peak of the $\chi$ spectrum is slightly broader and shifted towards the infrared. 

For the sake of completeness we also present the spectra obtained from the simulation with $m_{\chi}/m_{\phi}=1$. As one would expect from the variances in fig.~\ref{fig:vars}, the $\chi$ modes do not experience a significant growth, while the spectrum of $\phi$ develops the usual peak during the tachyonic oscillations phase. 
\begin{figure}[tbp]
\centering
\includegraphics[width=12cm]{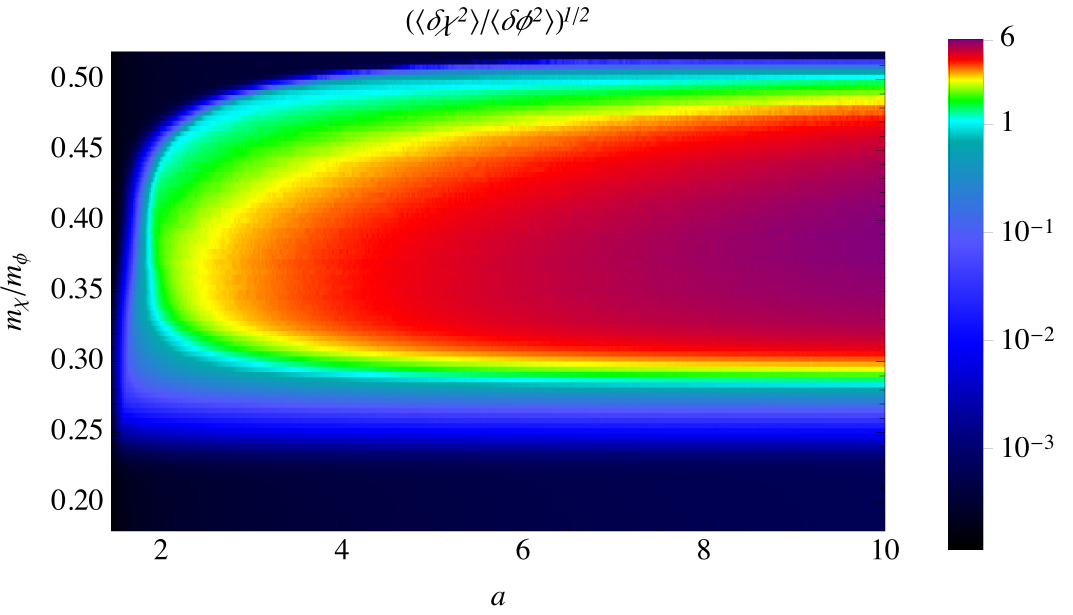}
\caption{The evolution of $\sqrt{\braket{\delta\chi^2}/\braket{\delta\phi^2}}$ as a function of $a$ for $v=10^{-2}m_{\rm Pl}$ and for different values of $m_{\chi}/m_{\phi}$. The plot was obtained from $94$ lattice simulations in two spatial dimensions with $1024$ points per dimension. We indeed find a broad resonance band for exactly the range of $m_\chi/m_\phi$ for which our Floquet analysis in section~\ref{sec:floquet} predicts the strongest enhancement (cf.\ fig.~\ref{fig:floquetChi}).}
\label{fig:ratios}
\end{figure}

\subsection{Band structure of the resonance}
To study the dependence of the resonance on the mass ratio $m_\chi/m_\phi$ in more detail, we also performed an extensive analysis by scanning over $0.18 \lesssim m_{\chi}/m_{\phi} \lesssim 0.52$, in steps of $\Delta\frac{m_{\chi}}{m_{\phi}} = 0.0036$.\footnote{This corresponds to scanning over $0.5\,m^{-1}_{\rm Pl} \leq 10^3\lambda \leq 1.43\,m^{-1}_{\rm Pl}$ in steps of $\Delta\lambda = 10^{-5}\,m^{-1}_{\rm Pl}$. } Note that this analysis consists of $94$ lattice simulations. We therefore halved the number of lattice points per dimension keeping the same infrared cut-off and checked that the results do not change significantly.

Fig.~\ref{fig:ratios} shows the evolution of $\sqrt{\braket{\delta\chi^2}/\braket{\delta\phi^2}}$ as a function of $a$ for different values of $m_{\chi}/m_{\phi}$. One can clearly see the resonance band which extends from $m_{\chi}/m_{\phi}\simeq0.25$ to $m_{\chi}/m_{\phi}\simeq0.5$. Compared to our Floquet analysis in section~\ref{sec:floquet}, this corresponds to the region of the lower resonance band in fig.~\ref{fig:floquetChi} with a Floquet exponent ${\rm Re}(\mu)/H\gtrsim 20$. On the other hand, we did not observe any amplification of $\chi$ for $m_{\chi}/m_{\phi}\simeq1$, where our analysis in section~\ref{sec:floquet} shows some indication for a much weaker resonance band (cf.\ fig.~\ref{fig:floquetChi}). Note moreover that the ratio $\sqrt{\braket{\delta\chi^2}/\braket{\delta\phi^2}}$ did not reach its final value by the end of our simulation. This can also be seen from fig.~\ref{fig:vars} where it is clearly visible that $\braket{\delta\chi^2}$ and $\braket{\delta\phi^2}$ redshift differently.

For $m_\chi/m_\phi \lesssim 0.2$, we do not observe any amplification of $\chi$, which confirms our interpretation of the amplification as a parametric resonance since it only occurs for values of the coupling within a resonance band.

Note that, as mentioned in section \ref{sec:lattice_initial_conditions}, we have assumed zero initial conditions for $\braket{\chi}$, which is justified e.g.\ in the presence of small supergravity corrections. If such corrections are absent, $\braket{\chi}$ can be non-zero initially, i.e.\ when the fields leave the quantum diffusion region close to the top of the hill. Interestingly, as we discuss in appendix \ref{sec:nonzerochi}, from analysing the quantum diffusion regime one can derive an upper bound on the initial values of $|\braket{\chi}|$ as a function of $m_\chi/m_\phi$. This bound implies that for $m_\chi/m_\phi \gtrsim 0.5$, i.e.\ in the upper part of and above the resonance band in fig.~\ref{fig:ratios}, one can neglect the effects of the non-zero initial conditions for $\braket{\chi}$. However, for smaller values of $m_\chi/m_\phi$, non-zero initial $\braket{\chi}$ within the derived limits can enhance $\sqrt{\braket{\delta\chi^2}/\braket{\delta\phi^2}}$ and thereby enlarge the resonance band somewhat down to smaller mass ratios.

\subsection{Interpretation as parametric resonance}

Figs.~\ref{fig:floquetChi}--\ref{fig:ratios} indicate that we indeed see a parametric resonance of $\chi$ from an inhomogeneous $\delta \phi$ background: the growth of $\chi$ is rapid and exponential, and it only occurs for specific $\lambda$ (or specific $m_\chi/m_\phi$ respectively) and $k$ within a resonance band, with no exponential growth occuring for smaller or larger values of $\lambda$. Also, the position of the resonance band coincides with the estimate derived from eq.~\eqref{eq:eomChi3}, which is mathematically equivalent to the equations of multi-field parametric resonance.

\section{Summary and conclusions}\label{sec:conclusions}

In this paper we have studied preheating after hilltop inflation in the presence of another scalar field $\chi$ that couples to the inflaton $\phi$. We saw that, after the inflaton has become inhomogeneous, $\chi$ can be amplified from its initial vacuum fluctuations up to amplitudes of the order of the inflaton fluctuations. This amplification is due to a parametric resonance of $\chi$ caused by the inhomogeneous inflaton field.

In order to understand the parametric dependence of the amplification, in section~\ref{sec:floquet} we carried out a semi-analytical generalized Floquet analysis for inhomogeneous background fields. For the model~\eqref{eq:potential} with $v=10^{-2}m_{\rm Pl}$, we found a strong resonance band around a mass ratio of  $m_{\chi}/m_{\phi}\lesssim 0.5$ and some indication for a much weaker band around $m_{\chi}/m_{\phi}\sim 1$, see fig.~\ref{fig:floquetChi}. 

Guided by the semi-analytical results, we ran lattice simulations of the model for $v=10^{-2}\mpl$ and interaction parameters within the resonance bands. The results are described in section~\ref{sec:num_analysis_lattice}. We confirmed the existence of the resonance band at $0.25 \lesssim m_{\chi}/m_{\phi}\lesssim 0.5$ for which the $\chi$ fluctuations can be amplified to amplitudes even exceeding those of the inflaton field, see fig~\ref{fig:ratios}. The time at which the amplification starts depends on the exact mass ratio and it typically is of the order of $1$ $e$-fold after the inflaton field has become inhomogeneous. We could not find any amplification corresponding to the indication for the possibility of a weaker band around $m_{\chi}/m_{\phi}\sim 1$ from the Floquet analysis.

Our results highlight the importance of studying preheating after hilltop inflation until well after non-linear dynamics start dominating the evolution: resonant production of $\chi$ can occur even after the inflaton field has become inhomogeneous. This effect has to be taken into account when calculating the initial particle abundances for the final stages of reheating, in order to accurately compute quantities like e.g.\ the reheat temperature, non-thermally produced relics and the baryon asymmetry from non-thermal leptogenesis.

\section*{Acknowledgements}
The authors would like to thank Vinzenz Maurer for useful discussions and help with optimizing our numerical tools. This work has been supported by the Swiss National Science Foundation.

\appendix
\section*{Appendix}

\section{Initial $\braket{\chi}$ from preinflation}\label{sec:nonzerochi}

In hilltop inflation, it is often assumed that the inflaton is the only field that is displaced from the vacuum during inflation, i.e.\ $\braket{\chi}=0$ for the hilltop model studied in this paper. In the context of preinflation with $\chi$ as the preinflaton, this can occur naturally in the presence of small supergravity corrections to the particle masses \cite{Antusch:2014qqa}.\footnote{These mass corrections are much smaller than all of the energy scales considered in this paper, so we would not expect an effect on our preheating analysis. However, they can dominate during the transition from preinflation in $\chi$ to hilltop inflation in $\phi$ and therefore determine the ratio $\braket{\chi}/\braket{\phi}$ at the beginning of hilltop inflation. For a large part of the parameter space, $\chi$ turns out to be exponentially suppressed, unless the supergravity corrections for $\phi$ are much larger than those for $\chi$.} However, $\braket{\chi}$ can also be non-negligible at the beginning of hilltop inflation, especially if no mass terms for $\chi$ and $\phi$ are generated from supergravity corrections.

The analysis in the previous sections, which assumed $\braket{\chi} = 0$, should therefore be interpreted as a conservative estimate on the abundance of $\chi$ fluctuations after preheating, showing that large $\chi$ fluctuations can be generated even if we start from vacuum fluctuations only. In this appendix, we want to discuss the initial values of $\braket{\chi}$ that arise from preinflation in the absence of supergravity corrections, how they depend on the mass ratio $m_\chi/m_\phi$, and how they affect preheating.

Since the fields are very homogeneous during preinflation and inflation, we will only deal with the homogeneous background fields $\phi = \braket{\phi}$ and $\chi = \braket{\chi}$ in this context, and throughout this appendix we drop the brackets to keep the notation simple.

\subsection{Quantum diffusion}
During preinflation in $\chi$, we have $\chi > \phi$, with $\phi \rightarrow 0$ due to the large effective mass $m_\phi^{\rm (eff)} = \lambda \chi^2$ for $\phi$ induced by the vacuum expectation value of $\chi$, whereas $\chi$ moves more slowly due to its shallower potential. Eventually, we end up close to the $\phi=0$ axis. However, for $\phi=0$, the gradient of the potential vanishes, and the classical slow-roll equations of motion predict static fields $\dot{\phi}=\dot{\chi}=0$. In this region, quantum fluctuations dominate over the classical field evolution, and we must include these fluctuations in the time evolution of the fields.

\subsubsection*{Quantum fluctuations versus classical field evolution}

For each light scalar field $\varphi$, we can estimate whether the classical evolution or the quantum fluctuations dominate the time evolution by comparing the classical change of the field value per Hubble time $t_\Hubble = 1/\Hubble$ to the growth of the quantum fluctuation amplitude:
\begin{align}
 \lvert \Delta \varphi_{\rm cl} \rvert \, &= \, \lvert \dot{\varphi}_{\rm cl} \times t_\Hubble \rvert \, \simeq \, \frac{\lvert V' \rvert}{3 \Hubble} \times \frac{1}{\Hubble} \, \simeq \, \mpl^2\frac{\lvert V' \rvert}{V} ~~\stackrel{?}{>}~~ \lvert \Delta \varphi_{\rm qu} \rvert \, \simeq \, \frac{\Hubble}{2\pi}.
 \label{eq:quantumDiffusionCondition}
\end{align}
Note that we assumed the quantum fluctuations for a massless scalar field in de Sitter space, i.e.\ $m_\varphi \ll \Hubble$ and $\Hubble =$ constant. This is a very good approximation for flat regions of the scalar potential, including the region around $\phi = \chi = 0$ that we want to study.\footnote{It is possible to perform the calculation with greater precision including the fields' potential and interactions as discussed in \cite{Finelli:2010sh}. However, in this appendix we are only interested in a rough estimate of the maximum value of $\chi$ at the end of inflation, for which such high precision is not necessary.}

When eq.~\eqref{eq:quantumDiffusionCondition} is satisfied, the evolution of $\varphi$ can be calculated from its classical equation of motion. Otherwise, $\varphi$ performs a random walk, moving around by $\Delta \varphi_{\rm qu} \sim \Hubble/2\pi$ per Hubble time.

\begin{figure}[ht]
\begin{center}
\hspace{27pt}\includegraphics[width=0.65\textwidth]{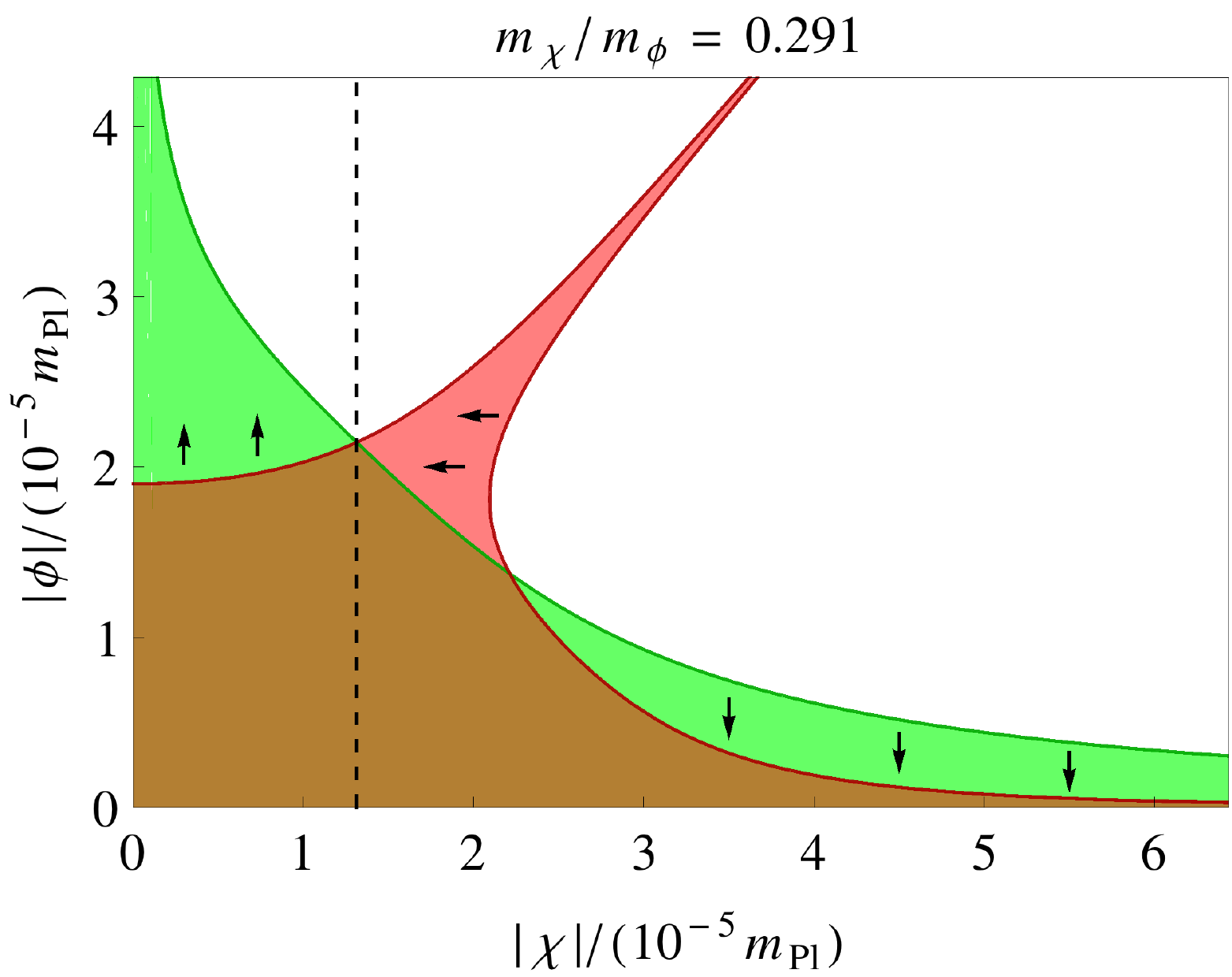}\newline\vspace{-23pt}
\end{center}
  \caption{Quantum diffusion region for $\lvert \chi \rvert$ and $\lvert \phi \rvert$ for $m_\chi/m_\phi = 0.291$. In the brown region to the lower left, both fields' evolution is dominated by quantum fluctuations and the fields perform a random walk. In the green regions, the field $\phi$ evolves according to its classical equation of motion whereas $\chi$ performs a random walk, and in the red region $\phi$ performs a random walk and $\chi$ is dominated by its classical equation of motion. The dashed black line marks a boundary between different regimes of the classical evolution: for $\lvert \chi \rvert > \chi_{\rm diff}^{(\rm max)} \simeq 1.3 \times 10^{-5}\mpl$, the classical equations push the fields back into the brown diffusion region when they randomly diffuse out of it, whereas for $\lvert \chi \rvert < \chi_{\rm diff}^{(\rm max)}$, the classical equations push $\lvert \phi \rvert$ towards large values away from the diffusion region. Therefore, the fields can only exit the brown diffusion region for small $\lvert \chi \rvert < \chi_{\rm diff}^{(\rm max)}$.}
  \label{fig:diffusionChi}
\end{figure}
\begin{figure}[ht]
\hspace{-10pt}
$\begin{array}{cc}
\includegraphics[width=0.5\textwidth]{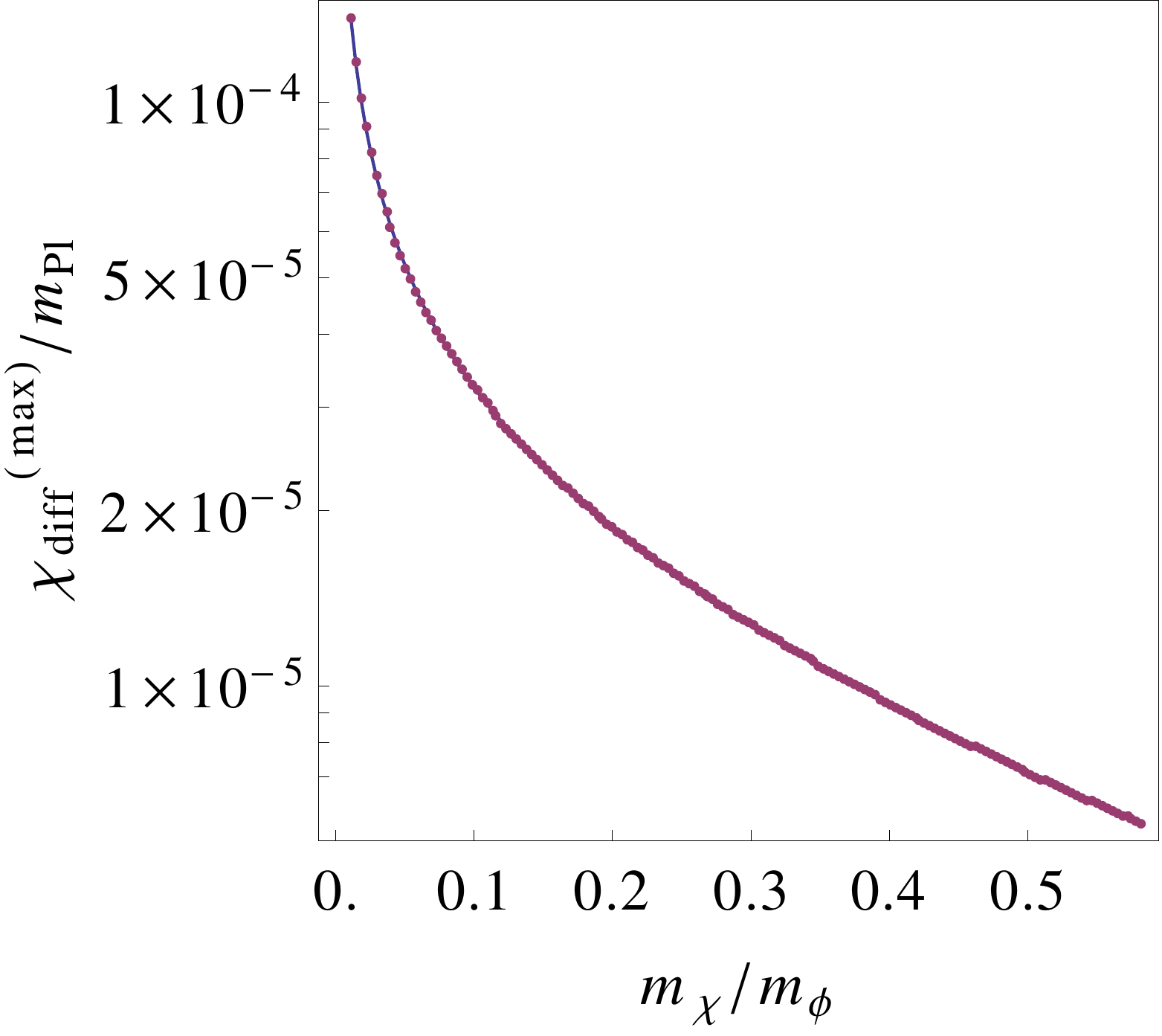} & \includegraphics[width=0.5\textwidth]{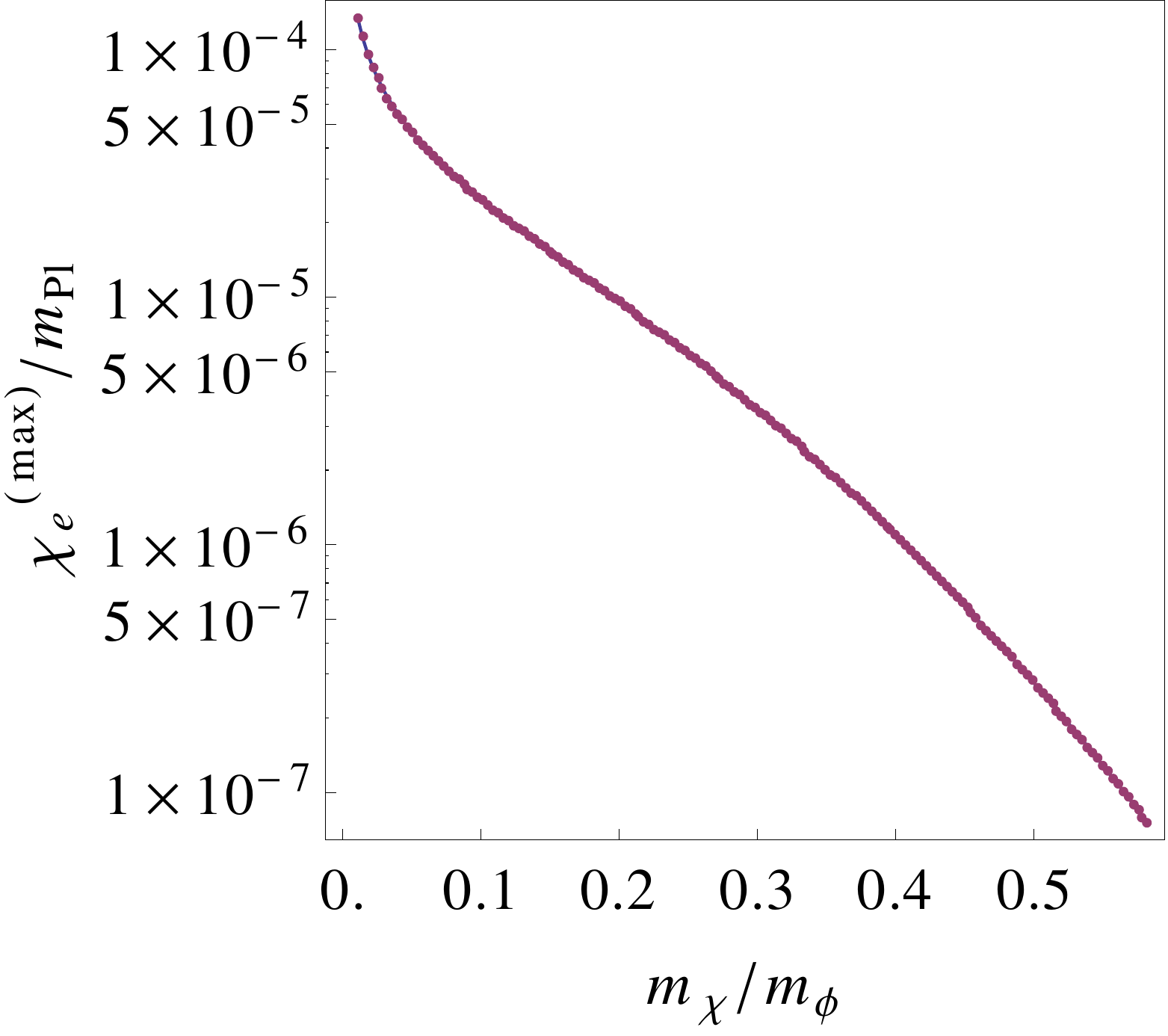}
\end{array}$
  \caption{Maximum values of $\chi$ at the diffusion region boundary (left) and close to the end of inflation when $\phi = 0.08v$ (right) as a function of $m_\chi/m_\phi$.}
  \label{fig:diffusionChiE}
\end{figure}

\subsubsection*{Diffusion region for $\phi$ and $\chi$}

We can now use the condition~\eqref{eq:quantumDiffusionCondition} to determine the ``diffusion region'' where $\phi$ and/or $\chi$ perform a random walk due to quantum fluctuations.

The diffusion region is plotted in fig.~\ref{fig:diffusionChi} for an example value $m_\chi/m_\phi = 0.291$. In the brown region, both $\phi$ and $\chi$ perform a random walk. In the green regions, the field $\phi$ evolves according to its classical equation of motion whereas $\chi$ performs a random walk, and in the red region $\phi$ performs a random walk and $\chi$ is dominated by its classical equation of motion.

To understand where we can leave the brown diffusion region, it is important to understand the evolution in the red and green regions:
\begin{enumerate}
 \item In the red region, $\chi$ moves classically and $\phi$ performs a random walk with $\Delta \phi \sim \Hubble/2\pi$ per Hubble time. The dominant effect is the classical movement of $\chi$, which drives $\chi$ to smaller values; in the red region, we thus move nearly horizontally to the left. Note that this means that when we exit from the brown to the red region, the field is immediately pushed back into the brown diffusion region.
 \item In the green regions, the dominant effect is the classical movement of $\phi$. For $\lvert \chi \rvert > \chi_{\rm diff}^{(\rm max)}$ (to the right of the dashed black line), the mass of $\phi$ is positive, and we move vertically downwards back into the brown diffusion region. For $\lvert \chi \rvert < \chi_{\rm diff}^{(\rm max)}$, the mass of $\phi$ is tachyonic, and we move vertically upwards away from the brown diffusion region.
\end{enumerate}
The combined effect is that the fields can only exit the brown diffusion region at $\lvert \chi \rvert < \chi_{\rm diff}^{(\rm max)}$. Where exactly they exit the diffusion region is the result of a random walk through the brown region starting at $\lvert \chi \rvert \gg \chi_{\rm diff}^{(\rm max)}$.

The value $\chi_{\rm diff}^{(\rm max)}$ can easily be calculated as the intersection between the diffusion boundaries of $\phi$ and $\chi$:
\begin{align}
 \text{At }(\phi_{\rm diff}^{(\rm max)}, \chi_{\rm diff}^{(\rm max)}): \quad~~ \frac{1}{V}\frac{\partial V}{\partial \chi} \, = \, -\frac{1}{V} \frac{\partial V}{\partial \phi} \, = \, \frac{\Hubble}{2\pi\mpl^2}.
\end{align}
The minus sign selects the upper left boundary of the diffusion region of $\phi$; this is the boundary beyond which $\phi$ is growing, whereas beyond the lower right boundary, the classical motion drives $\phi \rightarrow 0$.

\subsection{Dependence of initial $\braket{\chi}$ on $m_\chi/m_\phi$}

There are two effects that make $\chi$ dependent on the mass ratio $m_\chi/m_\phi$:
\begin{enumerate}
 \item The shape of the diffusion region depends on the coupling $\lambda \propto m_\chi/m_\phi$. For large $\lambda$, the potential for $\chi$ gets steeper; this makes the classical region larger and pushes the green and brown regions in fig.~\ref{fig:diffusionChi} to the left, towards smaller $\chi$, and thus reduces $\chi_{\rm diff}^{\rm(max)}$.
\item After the fields leave the diffusion region at some $\chi_{\rm diff}$, $\chi$ rolls along the potential gradient towards smaller $\chi$, with $\dot{\chi} \propto \lambda^2 \propto (m_\chi/m_\phi)^2$. Therefore, a larger mass ratio makes $\chi$ decay faster during inflation, reducing the value $\chi_{\rm e}$ at the end of inflation.\footnote{For our plots, we always show $\chi_{\rm e}$ at the time when $\phi = 0.08v$ which is the initial time for our lattice simulations.}
\end{enumerate}
Both of these effects make $\chi_{\rm diff}$ and especially $\chi_{\rm e}$ monotonously decreasing with $m_\chi/m_\phi$, see fig.~\ref{fig:diffusionChiE}. For this reason, a noticeable effect of $\chi_{\rm e}$ on preheating is only possible for small mass ratios $m_\chi/m_\phi \lesssim 0.5$. For larger $m_\chi/m_\phi$, $\chi_{\rm e}$ is generally subdominant compared to the vacuum quantum fluctuations $\delta \chi^{\rm(vac)}$.

\subsection{Effect of non-negligible initial $\braket{\chi}$ on preheating}
\label{sec:latticeNonzeroChi}

The basic effect of a large $\chi_{\rm e}$ is that $\chi$ fluctuations after preheating tend to end up larger than they would be for $\chi_{\rm e}=0$. We have confirmed this by a series of lattice simulations using $\chi_{\rm e}^{\rm(max)}$ (as shown in fig.~\ref{fig:diffusionChiE}) as the initial condition for $\braket{\chi}$ to estimate the maximum possible effect of an initial $\lvert \braket{\chi} \rvert$.

For $m_\chi/m_\phi \gtrsim 0.5$, the maximum possible $\chi_{\rm e}$ is so small that the results of section~\ref{sec:num_analysis_lattice} remain unchanged. For small $m_\chi/m_\phi$, the initial $\chi_{\rm e}$ extends the range of parameters for which $\delta\chi$ can be large after preheating towards smaller $m_\chi/m_\phi$. In that case, the results of the previous sections should be interpreted as a lower bound on the abundance of $\chi$ particles after preheating.

\section{Parametric resonance of $S$}\label{appendix:S}

In the supersymmetric version of the model, there is an additional scalar singlet field $S$ which is stabilized at $S=0$ during inflation if we assume homogeneous fields. However, the mass of $S$ around $\phi \sim v$ depends on $\phi$:
\begin{align}
 V_S \, = \, m_{\phi}^2 \left( \frac{\phi}{v} \right)^{10} \lvert S \rvert^2.
\end{align}
Due to this inflaton-dependent mass, the quantum fluctuations of $S$ can be amplified during preheating. Note that around the minimum, $m_S^2 = m_\phi^2$ holds independently of the model parameters.

\subsubsection*{Floquet analysis}

\begin{figure}[tbp]
\hspace{-15pt}
$\begin{array}{ccc}
\includegraphics[width=0.51\textwidth]{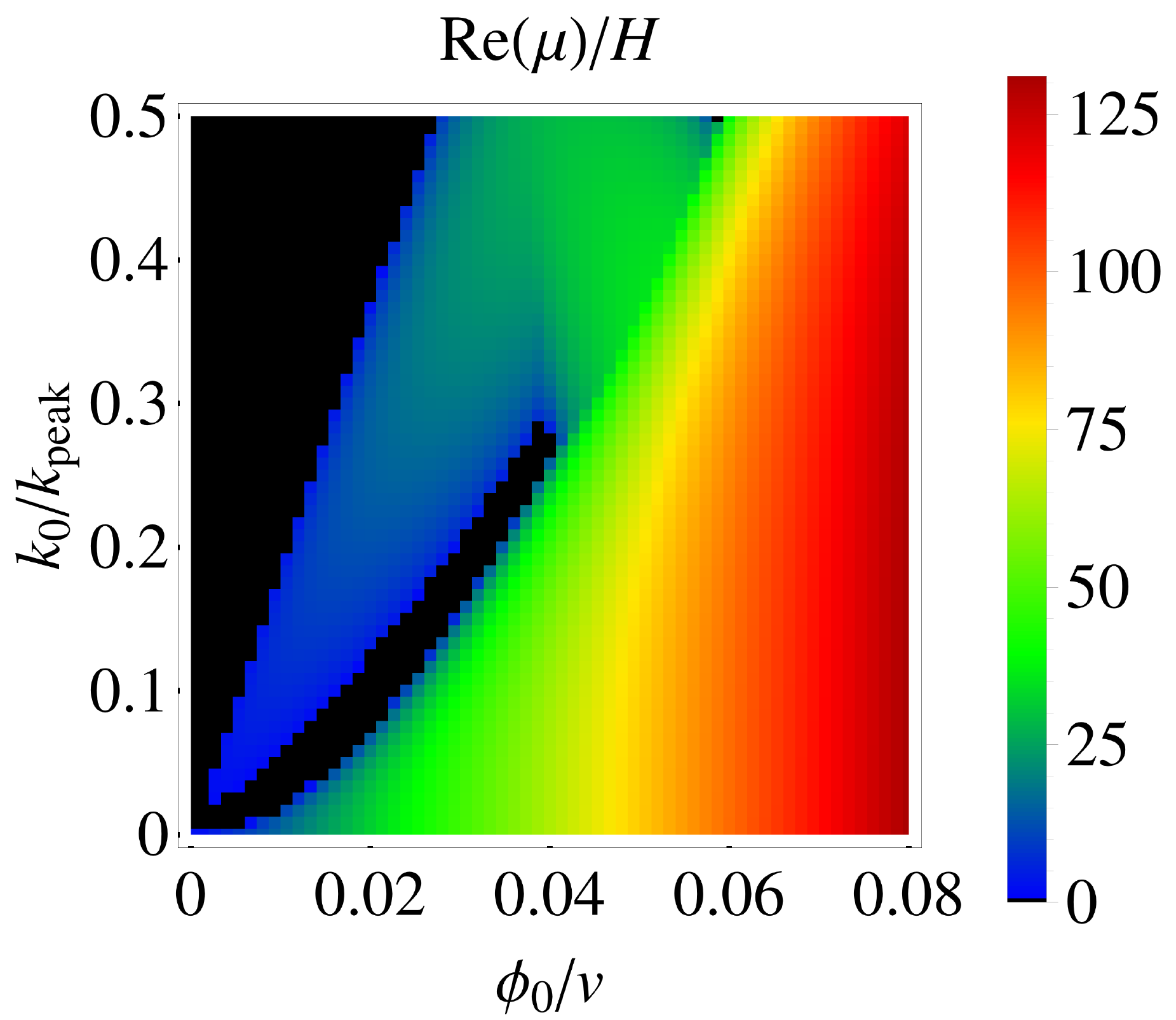} &
\includegraphics[width=0.51\textwidth]{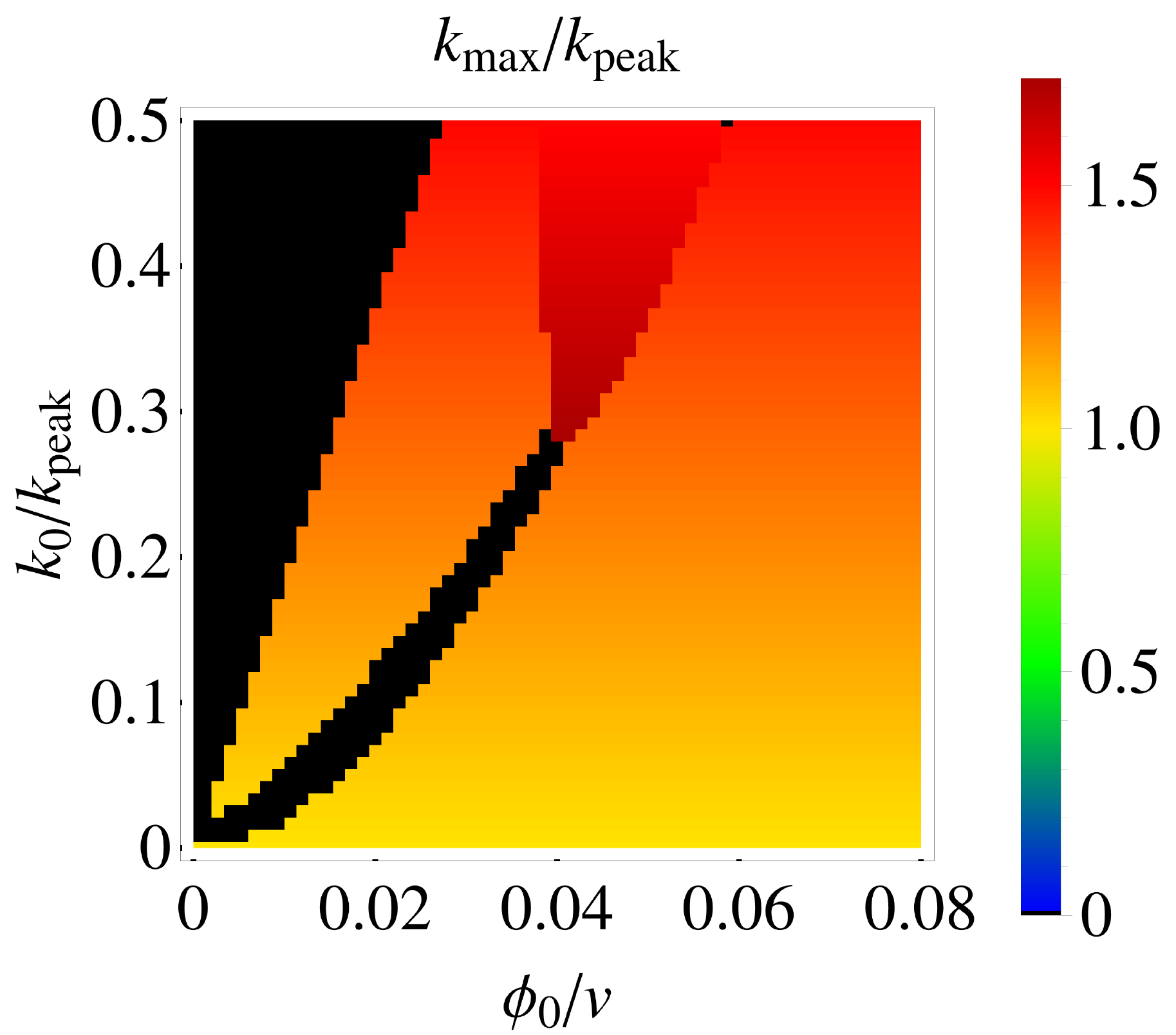}
\end{array}$
  \caption{Results of Floquet analysis for the perturbations of $S$ for $v=10^{-2} \mpl$, using a cutoff $k_{\rm cutoff} = 5k_{\rm peak}$. The left plot shows the largest Floquet exponent for a given $k_0$ and inflaton perturbation amplitude $\phi_0$, indicating how strongly the fastest-growing $\delta S$ perturbations are growing. Black regions indicate no growth (only oscillatory solutions), while in the coloured regions $\delta S$ grows exponentially. The right plot shows the wavenumber $k_{\rm max}$ around which the fastest-growing linear combination of $S_k$ is peaked. We find that strong growth occurs especially around $k \sim k_{\rm peak}$.}
  \label{fig:floquetS}
\end{figure}

Some preheating already takes place during the tachyonic oscillation phase (during which $\phi$ is mostly homogeneous), and the amplification during this phase can be calculated using the standard techniques.

However, even after preheating has made $\phi$ inhomogeneous, parametric resonance for $S$ can still occur. Assuming, as in section~\ref{sec:floquet}, that $\phi(\vec{x},t)$ can be approximated by a single plane wave with wave vector $\vec{k}_{\rm p}$, we find equations for the real and imaginary parts of $S$ analogous to eqs.~\eqref{eq:eomChi3}--\eqref{eq:U}, with the matrix $\mathcal{F}(t)$ replaced by
\begin{align}
 \mathcal{F}_{S}(t) \, = \, -m_\phi^2 \mathbb{1} - \begin{pmatrix}
 k_{N}^2 & f_S(t) & 0 & 0 & ...& 0\\
 f_S(t) & k_{N-1}^2 & f_S(t) & 0 & ...& 0\\
 0 & f_S(t) & k_{N-2}^2 & f_S(t) & ...& 0\\
 0 & 0 & f_S(t) & k_{N-3}^2 &... & 0\\
 ... & ... & ... & ... & ... & f_S(t) \\
 0 & 0 & 0 & 0 & f_S(t)  &k_{-N}^2\\
\end{pmatrix},
\end{align}
and
\begin{align}
 f_S(t) \, := \, \frac{5m_\phi^2 \phi_0}{v} \cos(\omega_\phi t).
\end{align}

Performing the Floquet analysis for these equations with a cutoff $k_{\rm cutoff} = 5k_{\rm peak}$, we find the Floquet exponents shown in fig.~\ref{fig:floquetS}. The main result is that the perturbations $\delta S$ are strongly amplified, especially for momenta $k \sim k_{\rm peak}$. Note that these Floquet exponents do not depend on $\lambda$, unlike those for $\delta \chi$, and $\delta S$ grows exponentially even for parameters for which $\delta \chi$ remains small.

\subsubsection*{Lattice results}
Here, we present two lattice simulations including the field $s=\sqrt{2}\,{\rm Re}[S]$. The corresponding scalar potential is given by eq.~\eqref{eq:scalar_potential_sugra}, keeping the $S$ dependent terms:
\begin{align}
V_{s} \, = \, \frac{m^2_{\phi}}{2}s^2\left(\frac{\phi}{v}\right)^{10} - 6\,\sqrt{2}\,\lambda s\chi^2\left(\frac{\phi}{v}\right)^6\,.
\end{align}
The configuration of the lattice as well as the initial conditions of $\phi$ and $\chi$ are the same as in section~\ref{sec:num_analysis_lattice} (cf.\ table~\ref{tab:ic_and_parameters}). The field $s$ is initialized with zero initial conditions, i.e.\ $\langle s\rangle_{\rm i} = \langle \dot{s}\rangle_{\rm i} = 0$.

\begin{figure}[ht]
\centering
\subfigure{\includegraphics[width=7.5cm]{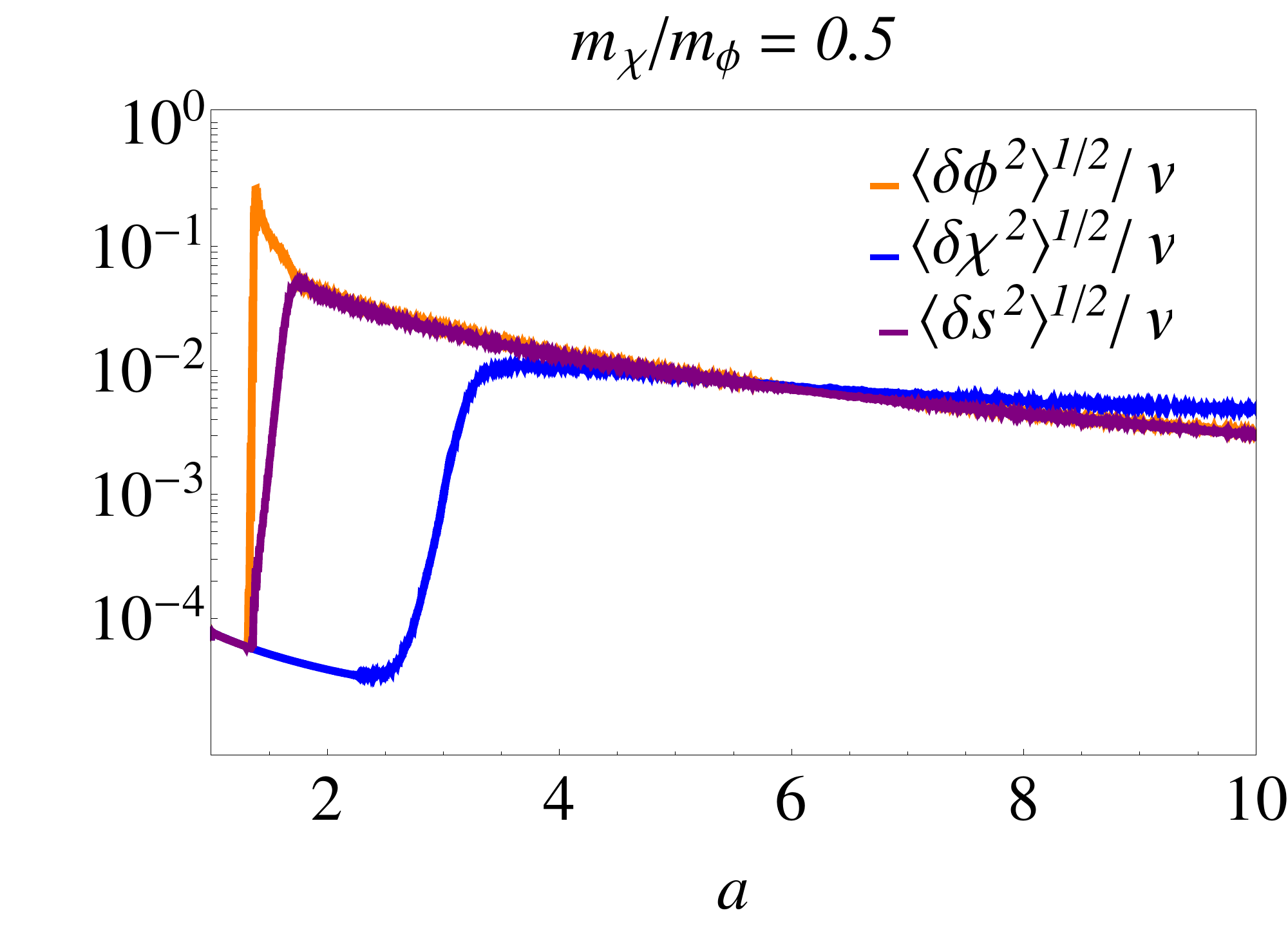}}
\hfill
\subfigure{\includegraphics[width=7.5cm]{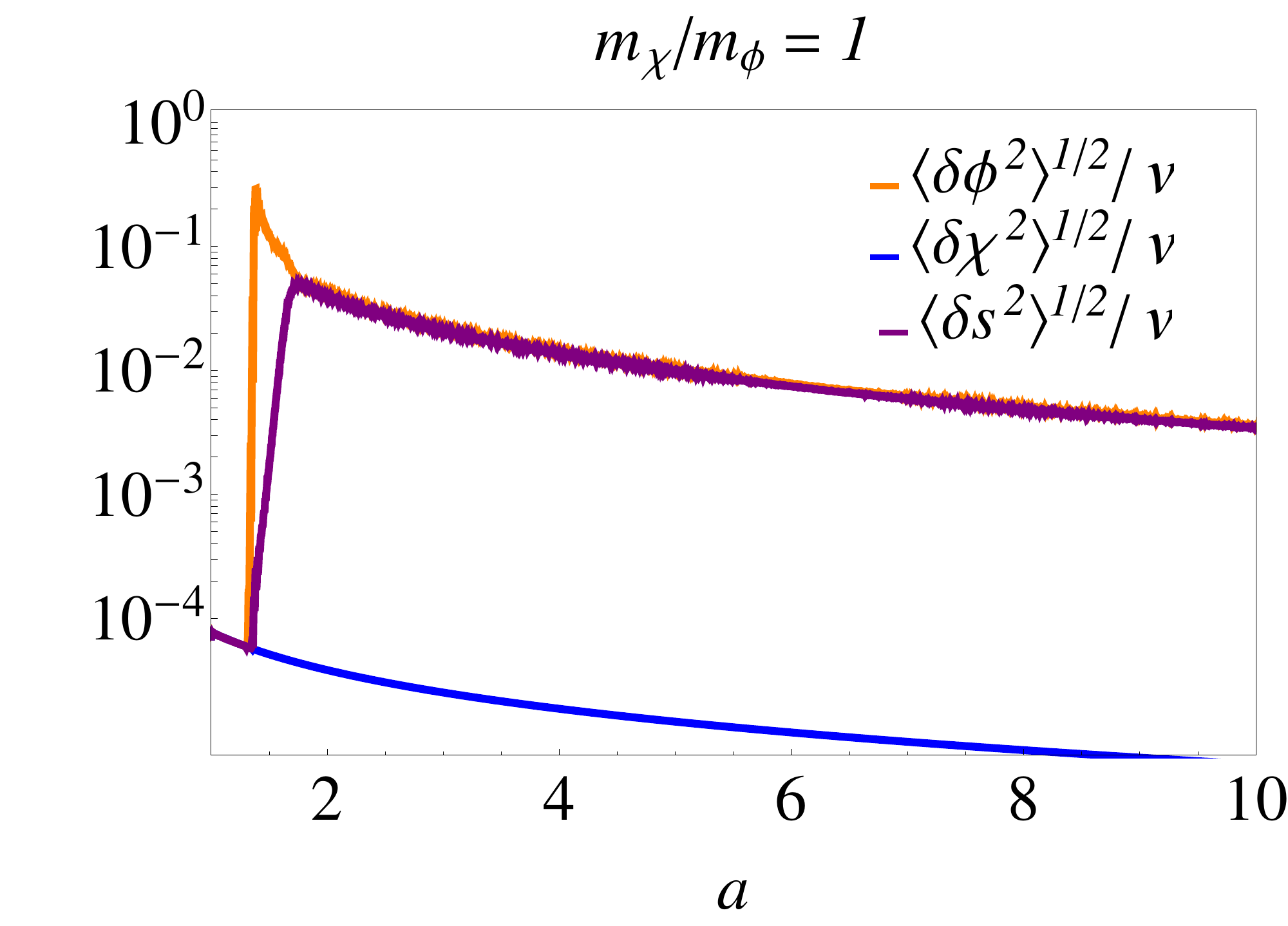}}
\caption{Results of lattice simulations including the field $s=\sqrt{2}\,{\rm Re}[S]$, for $v=10^{-2}m_{\rm Pl}$ and $m_{\chi}/m_{\phi}=0.5$ (left) and $m_{\chi}/m_{\phi}=1$ (right). For each mass ratio, we plot $\sqrt{\braket{\delta\phi^2}}/v$, $\sqrt{\braket{\delta\chi^2}}/v$ and $\sqrt{\braket{\delta s^2}}/v$ as a function of the scale factor $a$. One can see that in both cases, the variances of $\phi$ and $\chi$ evolve in the same way as in the simulations without $s$ (cf.\ fig.~\ref{fig:vars}). In both cases, the variance of $s$ experiences a resonant growth, starting during the homogeneous $\phi$ oscillations and continuing (with a noticeably different growth rate) when $\delta\phi$ has become non-linear, until $\sqrt{\braket{\delta s^2}}\sim\sqrt{\braket{\delta \phi^2}}$.}
\label{fig:vars_S}
\end{figure}

Fig.~\ref{fig:vars_S} shows the variances $\sqrt{\braket{\delta \phi^2}}/v$, $\sqrt{\braket{\delta \chi^2}}/v$ and $\sqrt{\braket{\delta s^2}}/v$ as a function of $a$, for $m_{\chi}/m_{\phi} = 0.5$ (left) and $m_{\chi}/m_{\phi} = 1$ (right). The evolution of $\phi$ and $\chi$ does not qualitatively change due to the presence of $s$ (see fig.~\ref{fig:vars}). On the other hand, the field $s$ experiences the same resonant amplification regardless of the value of $m_{\chi}/m_{\phi}$. This is reasonable, since $m_{s}=m_{\phi}$ independently of the model parameters. The resonance in $s$ differs from the one in $\chi$ in the sense that it starts earlier, during the first few oscillations when the inflaton is still well-described by the homogeneous field equations. When non-linearities start to dominate the evolution of $\phi$, the growth in $\sqrt{\braket{\delta s^2}}$ continues (with a different growth rate) until the variances and spectra of $s$ and $\phi$ become practically identical and subsequently track each other.

\subsubsection*{Summary of the effects of $S$ on preheating}

In summary, perturbations of $S$ are always strongly amplified by parametric resonance, starting during the tachyonic oscillation phase and continuing when $\delta \phi$ is inhomogeneous and peaked around $k_{\rm peak}$. Our lattice results indicate that there are some effects on the preheating of $\chi$, but the main qualitative result is unchanged: we observe strong growth of $\chi$ in the broad resonance band for $m_\chi/m_\phi \lesssim 0.5$, but not at larger mass ratios. In particular, even in the presence of $S$, we do not observe non-perturbative growth of $\chi$ at $m_\chi = m_\phi$.

\end{document}